\documentclass[12pt,preprint]{aastex}

\usepackage{amsmath}
\usepackage{chngcntr}
\usepackage{graphicx}
\usepackage{epstopdf}
\usepackage{natbib}
\usepackage{rotating}
\usepackage{lscape}
\usepackage{longtable}
\begin{document}

\title{Investigating the Cosmic-Ray Ionization Rate in the Galactic Diffuse Interstellar Medium through Observations of H$_3^+$}

\author{Nick Indriolo\altaffilmark{1,2},
%Geoffrey A. Blake\altaffilmark{3},
%Miwa Goto\altaffilmark{4},
%Tomonori Usuda\altaffilmark{5},
%Takeshi Oka\altaffilmark{6},
%T. R. Geballe\altaffilmark{7},
%Brian D. Fields\altaffilmark{1,8}
Benjamin J. McCall\altaffilmark{1,3}
}
\altaffiltext{1}{Department of Astronomy, University of Illinois at Urbana-Champaign, Urbana, IL 61801}
\altaffiltext{2}{Present address: Department of Physics and Astronomy, Johns Hopkins University, 3400 North Charles Street, Baltimore, MD 21218; indriolo@pha.jhu.edu}
%\altaffiltext{3}{Division of Geological and Planetary Sciences and Division of Chemistry and Chemical Engineering, MS 150-21, California Institute of Technology, Pasadena, CA 91125}
%\altaffiltext{4}{Max-Planck-Institut f\"{u}r Astronomie, K\"{o}nigstuhl 17, Heidelberg D-69117, Germany}
%\altaffiltext{5}{Subaru Telescope, 650 North A'ohoku Place, Hilo, HI 96720}
%\altaffiltext{6}{Department of Astronomy and Astrophysics and Department of Chemistry, University of Chicago, Chicago, IL 60637}
%\altaffiltext{7}{Gemini Observatory, 670 North A'ohoku Place, Hilo, HI 96720}
%\altaffiltext{8}{Department of Physics, University of Illinois at Urbana-Champaign, Urbana, IL 61801}
\altaffiltext{3}{Department of Chemistry, University of Illinois at Urbana-Champaign, Urbana, IL 61801}

\begin{abstract}

Observations of H$_3^+$ in the Galactic diffuse interstellar medium (ISM) have led to various surprising results, including the conclusion that the cosmic-ray ionization rate ($\zeta_2$) is about 1 order of magnitude larger than previously thought.  The present survey expands the sample of diffuse cloud sight lines with H$_3^+$ observations to 50, with detections in 21 of those.  Ionization rates inferred from these observations are in the range $(1.7\pm1.3)\times10^{-16}$ s$^{-1}<\zeta_2<(10.6\pm8.2)\times10^{-16}$ s$^{-1}$ with a mean value of $\zeta_2=(3.5^{+5.3}_{-3.0})\times10^{-16}$ s$^{-1}$.  Upper limits ($3\sigma$) derived from non-detections of H$_3^+$ are as low as $\zeta_2<0.4\times10^{-16}$ s$^{-1}$.  These low upper-limits, in combination with the wide range of inferred cosmic-ray ionization rates, indicate variations in $\zeta_2$ between different diffuse cloud sight lines.  A study of $\zeta_2$ versus $N_{\rm H}$ (total hydrogen column density) shows that the two parameters are not correlated for diffuse molecular cloud sight lines, but that the ionization rate decreases when $N_{\rm H}$ increases to values typical of dense molecular clouds.  Both the difference in ionization rates between diffuse and dense clouds and the variation of $\zeta_2$ among diffuse cloud sight lines are likely the result of particle propagation effects.  The lower ionization rate in dense clouds is due to the inability of low-energy (few MeV) protons to penetrate such regions, while the ionization rate in diffuse clouds is controlled by the proximity of the observed cloud to a site of particle acceleration.

%Calculations of the cosmic-ray ionization rate from theoretical cosmic-ray spectra require a large flux of low-energy (MeV) particles to reproduce values inferred from observations.  Given the relatively short range of low-energy cosmic rays --- those most efficient at ionization --- the proximity of a cloud to a site of particle acceleration may set its ionization rate.  Variations in $\zeta_2$ are thus likely due to variations in the cosmic-ray spectrum at low energies resulting from the effects of particle propagation.  To test this theory, H$_3^+$ was observed in sight lines passing through diffuse molecular clouds known to be interacting with the supernova remnant IC 443, a probable site of particle acceleration.  Where H$_3^+$ is detected, ionization rates of $\zeta_2=(20\pm10)\times10^{-16}$ s$^{-1}$ are inferred, higher than for any other diffuse cloud.  These results support both the concept that supernova remnants act as particle accelerators, and the hypothesis that propagation effects are responsible for causing spatial variations in the cosmic-ray spectrum and ionization rate.  Future observations of H$_3^+$ near other supernova remnants and in sight lines where complementary ionization tracers (OH$^+$, H$_2$O$^+$, H$_3$O$^+$) have been observed will further our understanding of the subject.

\end{abstract}


\section{INTRODUCTION} \label{section_intro}

The rich chemistry responsible for producing many of the small molecular species observed in the interstellar medium (ISM) is thought to be driven primarily by ion-molecule reactions \citep{watson1973,herbst1973}.  To initiate this reaction network, a source of ionization is required that operates in both diffuse molecular ($n\sim100$~cm$^{-3}$, $T\sim70$~K) and dense molecular ($n\sim10^4$~cm$^{-3}$, $T\sim30$~K) clouds.  As absorption by atomic hydrogen severely attenuates the flux of photons capable of ionizing many of the most abundant species in the molecular ISM (e.g., H$_2$, He, O), cosmic rays are the dominant ionization mechanism in such regions.  This makes the cosmic-ray ionization rate a vital parameter in modeling interstellar chemistry.

Several estimates of the cosmic-ray ionization rate have been made over the past 50 years.  Both theoretical calculations \citep[e.g.,][]{hayakawa1961,spitzer1968,webber1998} and observational inferences \citep[e.g.,][]{odonnell1974,black1977,black1978,hartquist1978,hartquist1978a,vandishoeck1986,federman1996,vandertak2000,mccall2003,indriolo2007,gerin2010,neufeld2010} produced a wide range of ionization rates, from a few times $10^{-18}$~s$^{-1}$ to a few times $10^{-15}$~s$^{-1}$.  However, many of the more recent values inferred for diffuse clouds have tended to be on the order of $10^{-16}$~s$^{-1}$.

Following its initial detection in the ISM \citep{geballe1996}, H$_3^+$ has widely been regarded as the most direct tracer of the cosmic-ray ionization rate \citep{dalgarno2006} due to a simple underlying chemistry.  Observations of H$_3^+$ in dense clouds \citep{mccall1999,kulesa2002,brittain2004,gibb2010}, diffuse clouds \citep{mccall1998,mccall2002,mccall2003,indriolo2007,indriolo2010b,crabtree2011}, and toward the Galactic Center \citep{geballe1999,goto2002,goto2008,goto2011,oka2005,geballe2010} can be and have been used to constrain the ionization rate in each of these environments.  Here we present the most comprehensive survey of H$_3^+$ in diffuse molecular cloud sight lines to date, expanding and improving upon our analysis from \citet{indriolo2007}.  

\section{H$_3^+$ CHEMISTRY} \label{section_H3+chemistry}

As mentioned above, the chemistry associated with H$_3^+$ in the ISM is rather simple.  H$_3^+$ is formed in a two-step process, beginning with the ionization of H$_2$ by cosmic rays,
\begin{equation}
{\rm H}_2 + {\rm CR}\rightarrow {\rm H}_2^+ + e^- + {\rm CR}'
\label{re_CR_H2}
\end{equation}
and quickly followed by a reaction of H$_2^+$ with H$_2$,
\begin{equation}
{\rm H}_2^+ + {\rm H}_2\rightarrow {\rm H}_3^+ + {\rm H}.
\label{re_H2_H2+}
\end{equation}
Some amount of H$_2^+$ will be destroyed by dissociative recombination with electrons,
\begin{equation}
{\rm H}_2^+ + e^-\rightarrow \mathrm{H + H},
\label{re_H2+_e}
\end{equation}
or charge transfer to atomic hydrogen,
\begin{equation}
{\rm H}_2^+ + {\rm H}\rightarrow {\rm H}_2 + {\rm H}^+,
\label{re_H2+_H}
\end{equation}
before it can form H$_3^+$, but these processes are slow compared to reaction (\ref{re_H2_H2+}).  Overall, reaction (\ref{re_CR_H2}) is the rate-limiting step (it is many orders of magnitude slower than reaction (\ref{re_H2_H2+})) and can be taken as the formation rate of H$_3^+$.  Photoionization of H$_2$ also occurs, but is negligible compared to ionization by cosmic rays. This is because the ultraviolet and soft X-ray photons capable of ionizing H$_2$ ($E>15.4$ eV) are severely attenuated in the outer layers of diffuse clouds by atomic hydrogen, which has a lower ionization potential (13.6 eV).  While higher energy photons (e.g., hard X-rays and $\gamma$-rays) can penetrate diffuse molecular clouds, the lower ionization cross sections at these energies and smaller flux of such photons make their contribution to the ionization rate minimal.  As a result, H$_3^+$ should be formed primarily through the ionization of H$_2$ by cosmic rays.

The primary destruction mechanisms for H$_3^+$ are dependent on the environment under consideration.  In diffuse molecular clouds, H$_3^+$ is predominantly destroyed via dissociative recombination with electrons,
\begin{equation}
{\rm H}_3^+ + e^-\rightarrow {\rm H}_2 + {\rm H~or~H + H + H}.
\label{re_H3+_e}
\end{equation}
%\begin{eqnarray}
%{\rm H}_3^+ + e^-&\rightarrow& {\rm H}_2 + {\rm H} \nonumber \\
%                 &\rightarrow&\mathrm{H + H + H}.
%\label{re_H3+_e}
%\end{eqnarray}
In dense clouds however, where the electron fraction is much lower, H$_3^+$ is destroyed by proton transfer to neutrals such as CO, O, and N$_2$:
\begin{equation}
{\rm H}_3^+ + {\rm CO} \rightarrow \mathrm{H_2 + HCO^+},
\label{re_H3+_CO_HCO+}
\end{equation}
\begin{equation}
{\rm H}_3^+ + {\rm CO} \rightarrow \mathrm{H_2 + HOC^+},
\label{re_H3+_CO_HOC+}
\end{equation}
%\begin{subequations}
%\begin{eqnarray}
%{\rm H}_3^+ + {\rm CO}&\rightarrow&\mathrm{H_2 + HCO^+}, \label{re_H3+_CO_HCO+}\\
%{\rm H}_3^+ + {\rm CO}&\rightarrow&\mathrm{H_2 + HOC^+}, \label{re_H3+_CO_HOC+}
%\end{eqnarray}
%\label{re_H3+_CO_sum}
%\end{subequations}
%\vspace{-0.35in}
\begin{equation}
{\rm H}_3^+ + {\rm O}\rightarrow{\rm H}_2 + {\rm OH}^+,
\label{re_H3+_O}
\end{equation}
\begin{equation}
{\rm H}_3^+ + {\rm N}_2\rightarrow{\rm H}_2 + {\rm HN}_2^+.
\label{re_H3+_N2}
\end{equation}
These ion-molecule reactions demonstrate how H$_3^+$ essentially drives the chemistry in the ISM by generating molecular ions which can then go on to form even more complex species.   Because the chemistry surrounding H$_3^+$ is so simple, and because its formation arises directly as a result of ionization by cosmic rays, observations of H$_3^+$ can be a powerful probe of the cosmic-ray ionization rate.

%%%%%%%%%%%%%%%%%%%%%%%%%%%%%%%%%%%%%%%%%%%%%%%%%%%%%%%%%%%%%%%%%%%%%%%%%%%%%%%
\section{METHODS} \label{section_methods}

\subsection{Observations}

In diffuse molecular clouds, conditions are such that only the 2 lowest energy levels of H$_3^+$---the $(J,K)$=$(1,1)$ {\it para} and $(1,0)$ {\it ortho} states---are expected to be significantly populated.  As a result, observations that probe transitions arising from these 2 states will trace the entire content of H$_3^+$ along a line of sight.  We targeted 3 such transitions---the $R(1,1)^u$, $R(1,0)$, and $R(1,1)^l$---which are observable at 3.668083~$\mu$m, 3.668516~$\mu$m, and 3.715479~$\mu$m, respectively.
%Five such ro-vibrational transitions are accessible in the infrared: the $R(1,1)^u$, $R(1,0)$, $R(1,1)^l$, $Q(1,1)$, and $Q(1,0)$.  %Transition properties, including wavelength and dipole moment, are given in Table \ref{table_trans_prop}.
%, and Figure \ref{fig_atmosphere} shows the atmospheric transmission spectrum near these transitions.  
As the weak H$_3^+$ lines are located near various atmospheric CH$_4$ and HDO lines, precise removal of the telluric lines is vital for the purpose of accurately determining H$_3^+$ column densities, and is facilitated by the observation of telluric standard stars.  Because the $R(1,1)^u$ and $R(1,0)$ transitions are easily observed simultaneously (being separated by about 4~\AA), and because together they probe the entire population of H$_3^+$, they are the most frequently observed transitions.

Although observations of H$_3^+$ have become more commonplace, as evidenced by the publication list in Section \ref{section_intro}, they are still rather difficult.  Given the transition dipole moments \citep[see, e.g.,][and references therein]{goto2002} and a typical relative abundance of about $10^{-7}$--$10^{-8}$ with respect to H$_2$, H$_3^+$ absorption lines in diffuse molecular clouds are usually about 1--2\% deep, and thus require a signal-to-noise ratio (S/N) greater than about 300 on the continuum to achieve $3\sigma$ detections.  As a result, background sources must have $L$-band (3.5~$\mu$m) magnitudes brighter than about $L=7.5$~mag to be feasible targets for 8~m (or $L=5.5$~mag for 4~m) class telescopes.  Ideally, these background sources should also be early-type stars (OB) such that the continuum is free of stellar absorption features.  A list of all background sources targeted in our study, along with select line-of-sight properties, is presented in Table \ref{table_targets}.  Additionally, to adequately sample the intrinsically narrow interstellar absorption lines requires high spectroscopic resolving power ($R\gtrsim20,000)$.

Observations used in this study were taken between 1998 and 2010 with a variety of instrument/telescope combinations, including CGS4 \citep{mountain1990} at UKIRT, NIRSPEC \citep{mclean1998} at Keck, {\it Phoenix} \citep{hinkle2003} at both KPNO and Gemini South, and CRIRES \citep{kaufl2004} at VLT.  A log of all H$_3^+$ observations in diffuse molecular clouds utilized herein---including those previously published---can be found in Table 2.4 of \citet{indriolo2011}.  For specifics regarding instrument configurations, observing methods and the like, the reader is referred to the following sources: \citet{mccall1998,mccall2002,mccall2003,mccall2001,geballe1999,indriolo2007,indriolo2010b,indriolo2011,crabtree2011}.

\subsection{Data Reduction}

Beginning with raw data frames, our reduction procedure employed standard IRAF\footnote{http://iraf.noao.edu/} routines commonly used in spectroscopic data reduction.  Upon extracting one-dimensional spectra, data were tranferred to Igor Pro\footnote{http://www.wavemetrics.com/} where we have macros written to complete the reduction \citep{mccall2001}.  Spectra published in \citet{indriolo2007} were reprocessed here using a method that better accounts for bad pixels in the CGS4 array.  A full description of the new reduction procedure used for CGS4 data---and all of our reduction procedures applying to H$_3^+$ data in general---is presented in \citet{indriolo2011}.

%%%%%%%%%%%%%%%%%%%%%%%%%%%%%%%%%%%%%%%%%%%%%%%%%%%%%%%%%%%%%%%%%%%%%%%%%%%%%%%%%%%
\section{UPDATED RESULTS} \label{chapter_H3Panalysis}

\begin{figure}
\centering
\includegraphics[width=3.5in]{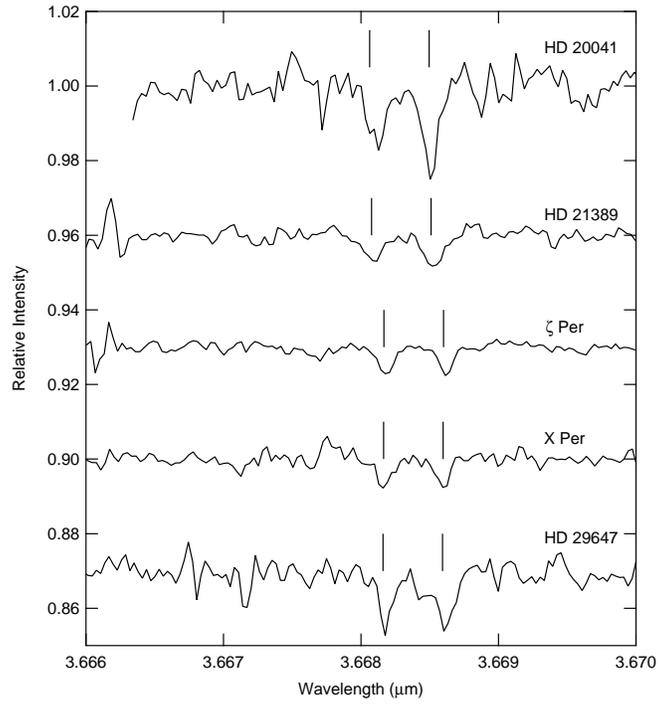}
\caption{Spectra targeting the $R(1,1)^u$ and $R(1,0)$ lines of H$_3^+$ in the sight lines toward HD 20041, HD 21389, $\zeta$ Per, X Per, and HD 29647.  Vertical lines mark the expected positions of absorption lines given gas velocities along each line of sight.  Gaps in spectra are where the removal of atmospheric features was particularly poor.  Observations were made with CGS4 at UKIRT.}
\label{fig_ukirt_det1}
\end{figure}

\begin{figure}
\centering
\includegraphics[width=3.5in]{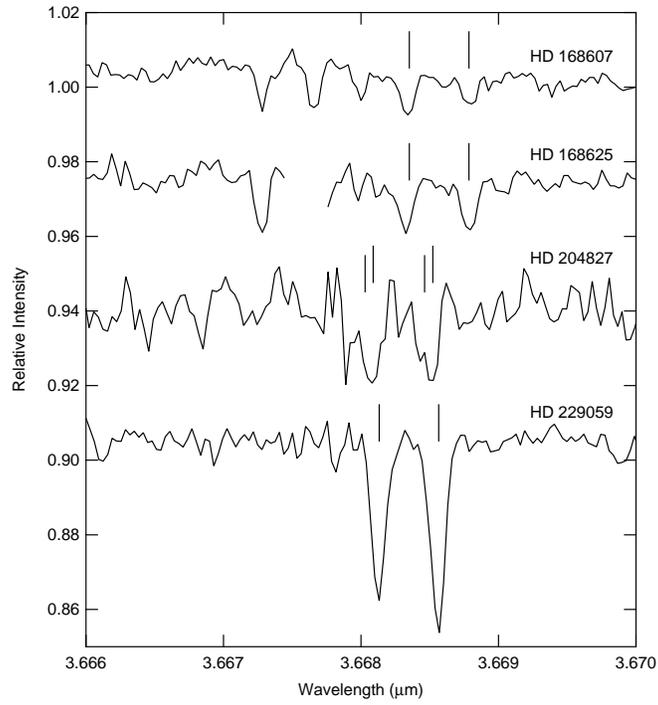}
\caption{Same as Figure \ref{fig_ukirt_det1} except for the sight lines toward HD 168607, HD 168625, HD 204827, and HD 229059.}
\label{fig_ukirt_det2}
\end{figure}

\begin{figure}
\centering
\includegraphics[width=3.5in]{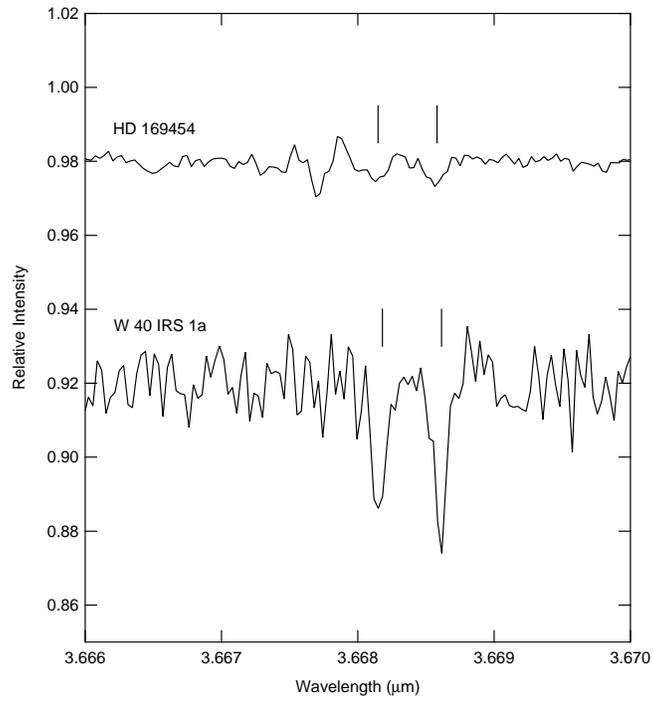}
\caption{Same as Figure \ref{fig_ukirt_det1} except for the sight lines toward HD 169454 and W 40 IRS 1A.}
\label{fig_ukirt_r10half}
\end{figure}

\begin{figure}
\centering
\includegraphics[width=3.5in]{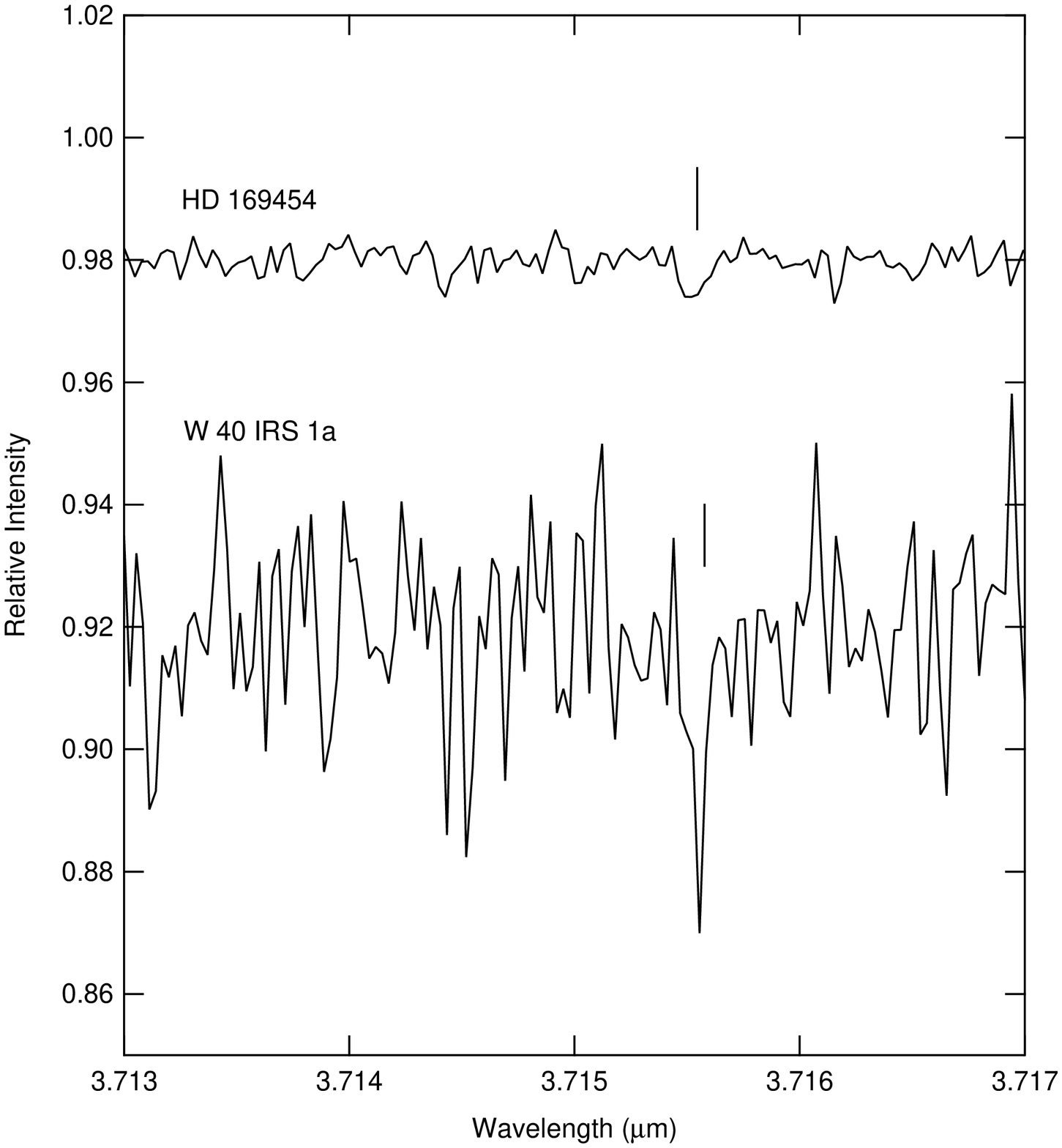}
\caption{Same as Figure \ref{fig_ukirt_r10half} except targeting the $R(1,1)^l$ transition.}
\label{fig_ukirt_r11lhalf}
\end{figure}

\begin{figure}
\centering
\includegraphics[width=3.5in]{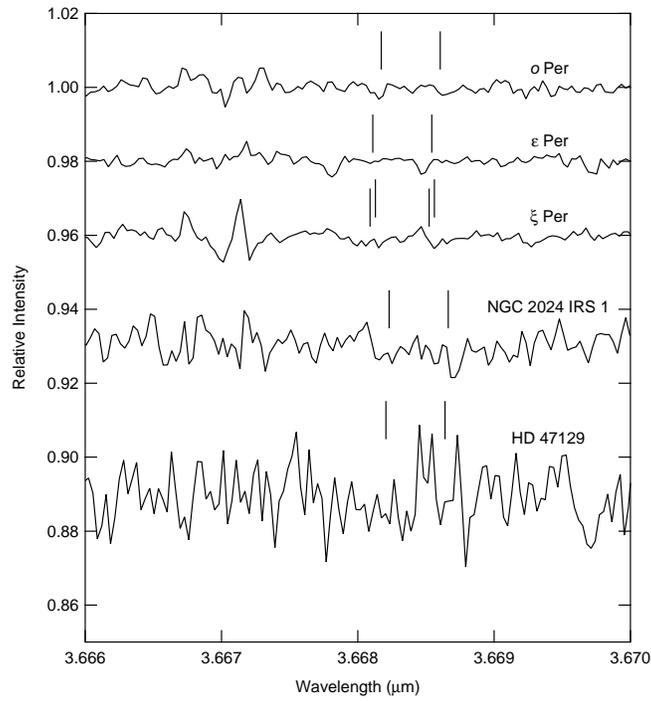}
\caption{Same as Figure \ref{fig_ukirt_det1} except for the sight lines toward $o$ Per, $\epsilon$ Per, $\xi$ Per, NGC 2024 IRS 1, and HD 47129.}
\label{fig_ukirt_nd1}
\end{figure}

\begin{figure}
\centering
\includegraphics[width=3.5in]{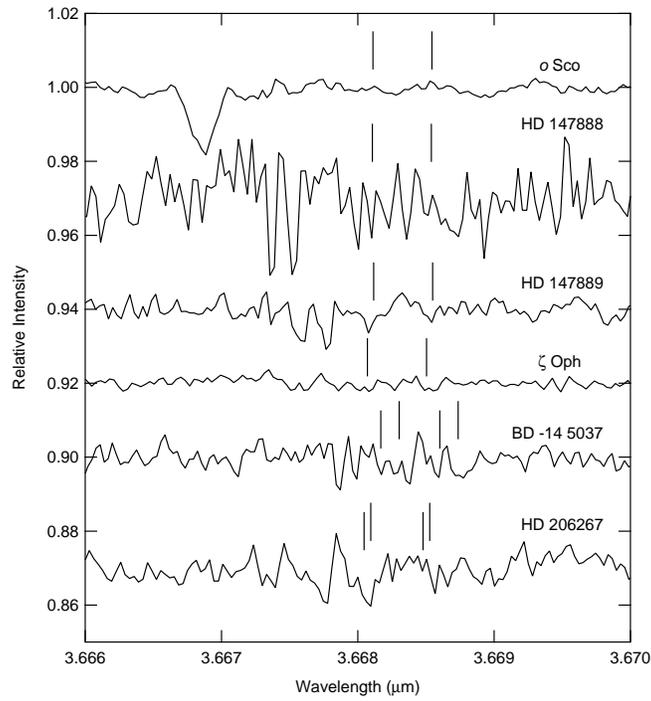}
\caption{Same as Figure \ref{fig_ukirt_det1} except for the sight lines toward $o$ Sco, HD 147888, HD 147889, $\zeta$ Oph, BD -14 5037, and HD 206267.}
\label{fig_ukirt_nd2}
\end{figure}

\begin{figure}
\centering
\includegraphics[width=3.5in]{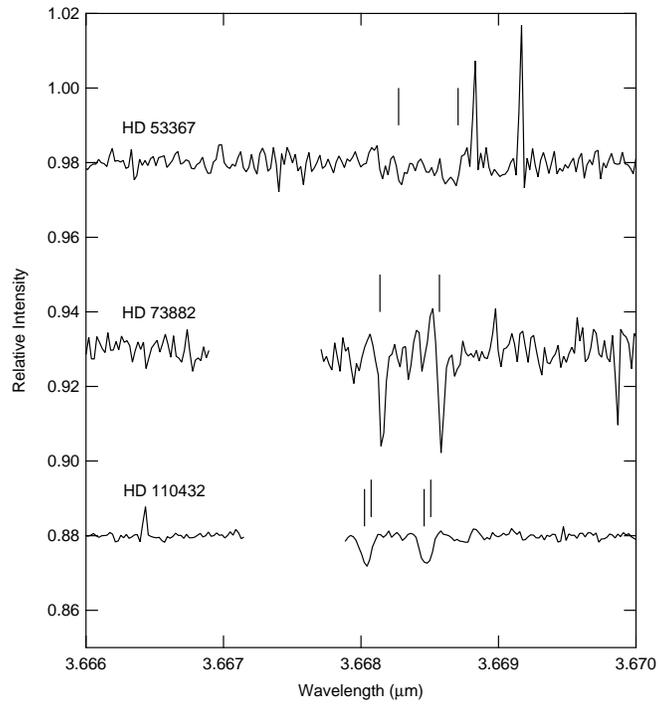}
\caption{Same as Figure \ref{fig_ukirt_det1} except for the sight lines toward HD 53367, HD 73882, and HD 110432.  Observations were made with CRIRES at VLT.}
\label{fig_vlt_r10half}
\end{figure}

\begin{figure}
\centering
\includegraphics[width=3.5in]{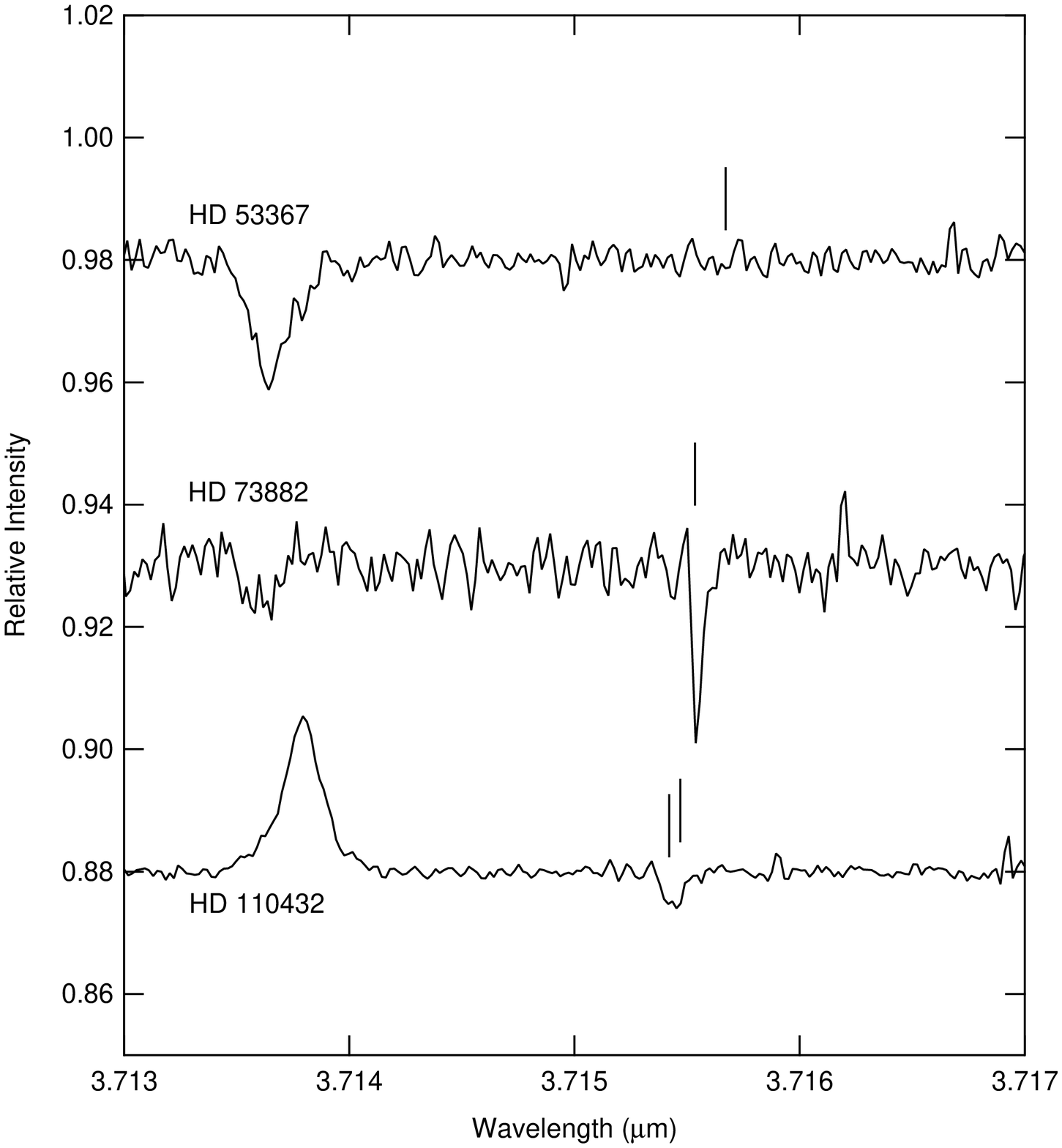}
\caption{Same as Figure \ref{fig_vlt_r10half} except targeting the $R(1,1)^l$ transition.}
\label{fig_vlt_r11lhalf}
\end{figure}

\begin{figure}
\centering
\includegraphics[width=3.5in]{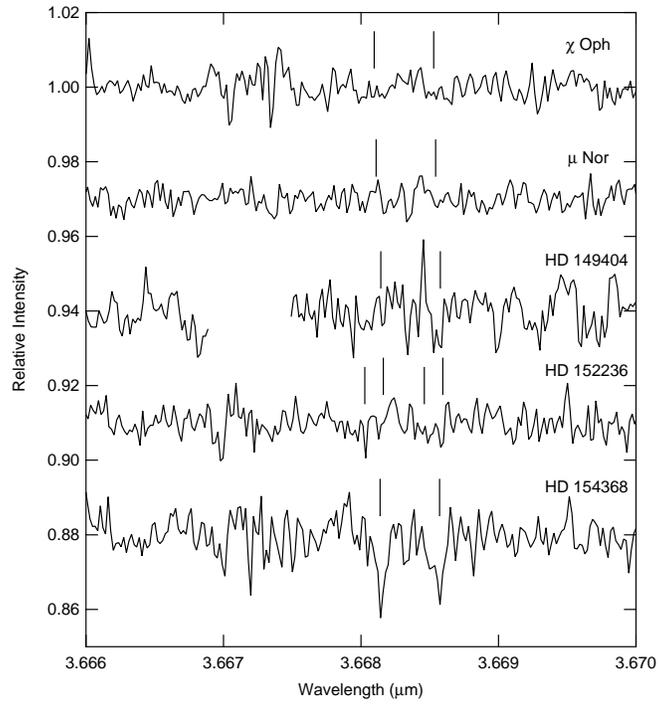}
\caption{Same as Figure \ref{fig_ukirt_det1} except for the sight lines toward $\chi$ Oph, $\mu$ Nor, HD 149404, HD 152236, and HD 154368.  Observations were made with {\it Phoenix} at Gemini South.}
\label{fig_gems}
\end{figure}

\begin{figure}
\centering
\includegraphics[width=3.5in]{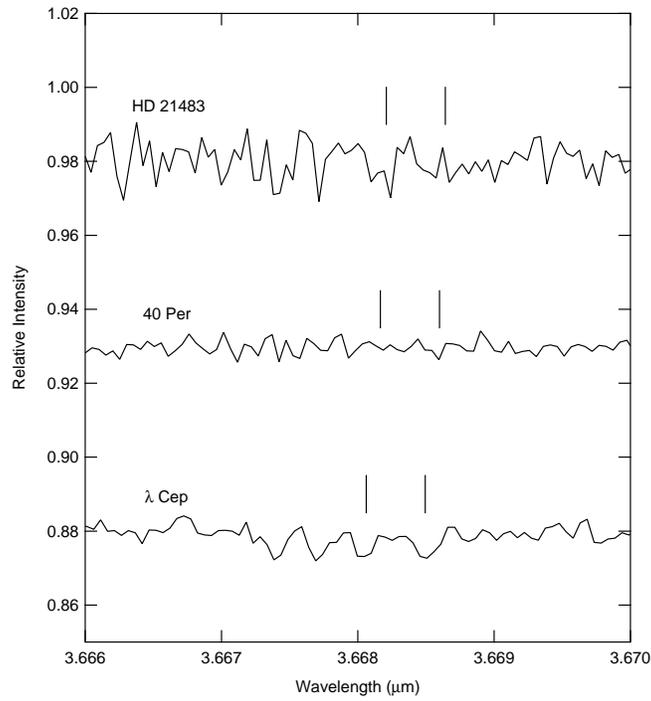}
\caption{Same as Figure \ref{fig_ukirt_det1} except for the sight lines toward HD~21483, 40~Per, and $\lambda$~Cep.  Spectra combined observations using CGS4 at UKIRT and NIRSPEC at Keck.}
\label{fig_keckukirt}
\end{figure}

\begin{figure}
\centering
\includegraphics[width=3.5in]{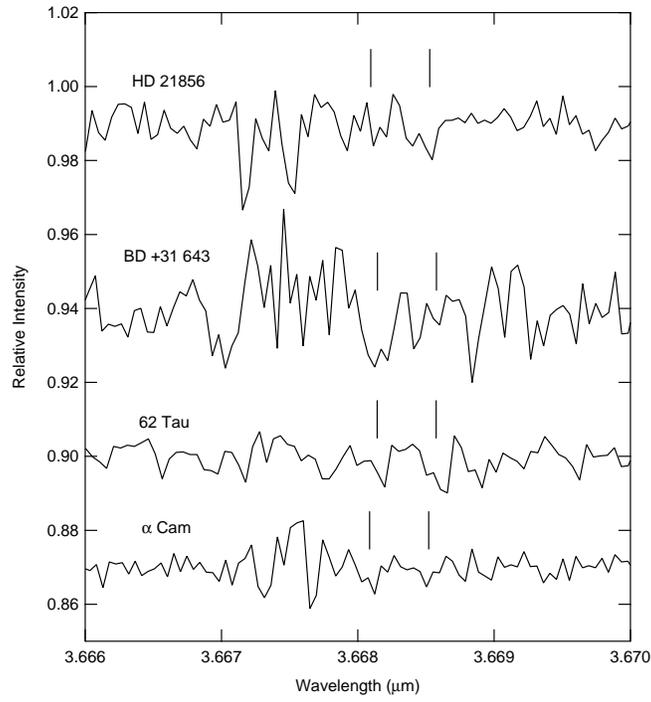}
\caption{Same as Figure \ref{fig_ukirt_det1} except for the sight lines toward HD~21856, BD~+31~643, 62~Tau, and $\alpha$~Cam.  Observations were made with NIRSPEC at Keck.}
\label{fig_keck1}
\end{figure}

\begin{figure}
\centering
\includegraphics[width=3.5in]{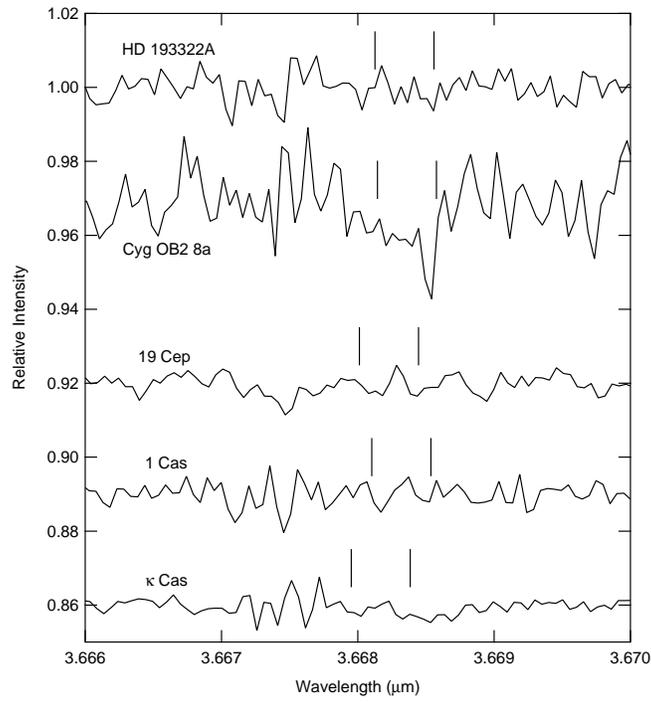}
\caption{Same as Figure \ref{fig_keck1} except for the sight lines toward HD~193322A, Cyg~OB2~8A, 19~Cep, 1~Cas, and $\kappa$~Cas.}
\label{fig_keck2}
\end{figure}

\begin{figure}
\centering
\includegraphics[width=3.5in]{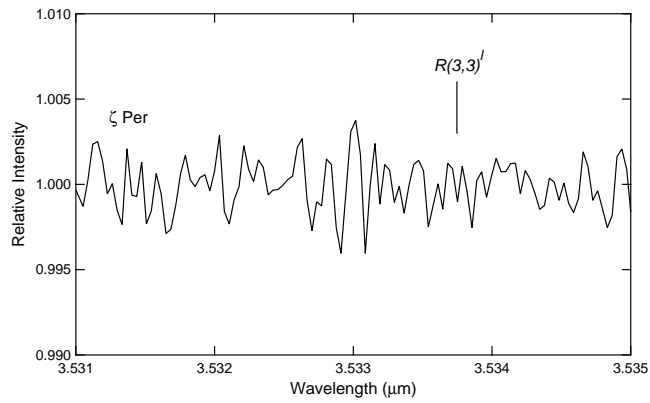}
\caption[GCS4 spectrum targeting the $R(3,3)^l$ line of H$_3^+$ toward $\zeta$~Per]{Spectrum covering the $R(3,3)^l$ line of H$_3^+$ toward $\zeta$~Per taken with CGS4 at UKIRT.  The vertical line marks the expected position of the absorption line.  Absorption from the (3,3) metastable state is not expected in the diffuse ISM, and thus far has only been detected in the Galactic Center \citep{goto2002,goto2008,goto2011,oka2005,geballe2010}.}
\label{fig_ukirt_r33l}
\end{figure}

Fully reduced spectra are presented in Figures \ref{fig_ukirt_det1}--\ref{fig_ukirt_r33l} .  Of the 42 sight lines presented, 15 show H$_3^+$ absorption lines, while 27 do not.  Combining these results with the spectra presented in \citet{mccall2002}, our entire survey covers 50 diffuse molecular cloud sight lines, with H$_3^+$ detections in 21 of those. 

As mentioned above, all of the spectra presented in \citet{indriolo2007} were re-processed for this study.  In most cases the differences between the resulting spectra are minimal, but spectra presented herein, along with any results derived from them, should be regarded as superseding any previously published results.  There are 4 sight lines in particular where further discussion of new versus old spectra is warranted: BD -14 5037, HD 154368, $\zeta$~Per, and X~Per.  In \citet{indriolo2007} we reported a detection of H$_3^+$ toward BD -14 5037---albeit marginal at best---but with the updated reduction procedure this sight line is now considered to be a non-detection (see Figure \ref{fig_ukirt_nd2} herein versus Figure 3 from \citet{indriolo2007}).  Spectra toward HD 154368, $\zeta$~Per, and X~Per were reported in \citet{crabtree2011}, but we later realized that an important step in the reduction process was omitted for these 3 sight linesß; spectra taken at the A and B nod positions had been combined based solely on pixel number without accounting for any potential shift in the dispersion direction.  For both HD 154368 and $\zeta$~Per the effects of re-processing the data were minimal, and newly derived column densities and {\it para} fractions agree with those reported in \citet{crabtree2011} within uncertainties.  For X~Per though, individual spectra had relatively low S/N and a large number of frames were shifted in the dispersion direction, thus the new reduction is significantly different from that reported in \citet{crabtree2011}.  By using the new X~Per results (see Table \ref{table_lineparam}) in the analysis presented by \citet{crabtree2011}, the conclusions in that study are actually strengthened, as the erroneous X~Per data used therein consistently made that sight line the only outlier in the reported trends.  

\section{ANALYSIS}

\subsection{Extraction of Column Densities and Upper Limits} \label{section_extract}

Absorption features due to H$_3^+$ were fit with Gaussian functions in order to determine equivalent widths, velocity FWHM, and interstellar gas velocities.  The fitting procedure uses the functional form of a Gaussian where the area (as opposed to amplitude) is a free parameter, and includes a fit to the continuum level; i.e.,
\begin{equation}
I=I_0-\frac{A}{w\sqrt{\pi}}\exp{\left[-\left(\frac{\lambda-\lambda_0}{w}\right)^2\right]},
%I=I_0-\frac{A_1}{w_1\sqrt{\pi}}\exp{\left[-\left(\frac{\lambda-\lambda_1}{w_1}\right)^2\right]},
%-\frac{A_2}{w_2\sqrt{\pi}}\exp{\left[-\left(\frac{\lambda-\lambda_2}{w_2}\right)^2\right]}.
\label{eq_fitgaussian}
\end{equation}
and was developed in Igor Pro.  The free parameters here are the continuum level, $I_0$, the area of the Gaussian, $A$, the central wavelength of the Gaussian, $\lambda_0$, and the line width, $w$ (n.b., the width used here is related to the ``standard'' Gaussian width, $\sigma$, by $w^2=2\sigma^2$).  As the continuum level has already been normalized, an area determined using this fit is by definition an equivalent width, $W_{\lambda}$.  In the case of the $R(1,1)^u$ and $R(1,0)$ lines, both absorption features are fit simultaneously and a single best-fit continuum level is found.  Uncertainties on the equivalent widths $(\delta W_{\lambda})$ and continuum level $(\delta I)$ --- both at the $1\sigma$ level --- are output by the fitting process.  To estimate the systematic uncertainties due to continuum placement, the continuum level was forced to $I_0+\delta I$ and $I_0-\delta I$ and the absorption lines re-fit.  Variations in the equivalent widths due to this shift are small compared to those reported by the fitting procedure and so have been ignored (i.e., $\sigma(W_{\lambda})=\delta W_{\lambda}$).  Assuming optically thin absorption lines and taking transition dipole moments and wavelengths from \citet[][and references therein]{goto2002}, column densities are derived from equivalent widths using the standard relation:
\begin{equation}
N(J,K)=\left(\frac{3hc}{8\pi^3}\right)\frac{W_{\lambda}}{\lambda}\frac{1}{|\mu|^2},
\label{eq_column}
\end{equation}
where $N(J,K)$ is the column density in the state from which the transition arises, $h$ is Planck's constant, $c$ is the speed of light, $\lambda$ is the transition wavelength, and $|\mu|^2$ is the square of the transition dipole moment.

In cases where H$_3^+$ absorption lines are not detected, upper limits to the equivalent width are computed as
\begin{equation}
W_{\lambda}<\sigma\lambda_{\rm pix}\sqrt{{\cal N}_{\rm pix}},
\label{eq_eqwid_ul}
\end{equation}
where $\sigma$ is the standard deviation on the continuum level near the expected H$_3^+$ lines, $\lambda_{\rm pix}$ is the wavelength coverage per pixel, and ${\cal N}_{\rm pix}$ is the number of pixels expected in an absorption feature.  Upper limits to the column density are then determined via equation (\ref{eq_column}).  Column densities, equivalent widths, velocity FWHM, and interstellar gas velocities for all of the sight lines presented in Figures \ref{fig_ukirt_det1}--\ref{fig_ukirt_r33l} are reported in Table \ref{table_lineparam}.

\subsection{Determination of the Cosmic-Ray Ionization Rate} \label{section_inferzeta}

As discussed in Section \ref{section_H3+chemistry}, the chemistry associated with H$_3^+$ in diffuse molecular clouds is rather simple. Assuming steady-state conditions, the formation and destruction rates of H$_3^+$ can be equated.  In the simplified version of diffuse cloud chemistry --- where destruction occurs only via dissociative recombination with electrons --- the result is
\begin{equation}
\zeta_{2}n({\rm H}_2)=k_en({\rm H}_3^+)n_e,
\label{eq_H3+_steadystate}
\end{equation}
where $\zeta_2$ is the ionization rate of H$_2$, $n$'s are number densities, and $k_e$ is the H$_3^+$-electron recombination rate coefficient (see Table \ref{table_H3+ratecoeff}).  Substituting the electron fraction (defined as $x_e\equiv n_e/n_{\rm H}$, where $n_{\rm H}\equiv n({\rm H})+2n({\rm H}_2)$) into equation (\ref{eq_H3+_steadystate}) and solving for the ionization rate gives
\begin{equation}
\zeta_2=k_{e}x_{e}n_{\rm H}\frac{n({\rm H}_3^+)}{n({\rm H}_2)}.
\label{eqzeta2numden}
\end{equation}
Observations cannot measure changes in these parameters along a line-of-sight, so we assume a uniform cloud with path length $L$ and constant $x_e$, $k_e$, $n_{\rm H}$, and $n({\rm H}_3^+)/n({\rm H}_2)$.  In this case, the H$_3^+$ and H$_2$ densities can by definition be replaced with $N({\rm H}_3^+)/L$ and $N({\rm H}_2)/L$, respectively, such that
\begin{equation}
\zeta_2=k_{e}x_{e}n_{\rm H}\frac{N({\rm H}_3^+)}{N({\rm H}_2)}
\label{eqzeta2}
\end{equation}
gives the cosmic-ray ionization rate in a diffuse molecular cloud.

%gives the average cosmic-ray ionization rate in the portion of a sight line that passes through a diffuse molecular cloud.

%Although it would be desirable to trace the ionization rate as a function of position throughout the cloud, variations in density along the line of sight cannot be determined via observations.  Instead, the average ionization rate in a cloud is inferred by using average number densities.  By definition, $\langle n({\rm H}_3^+)\rangle$ and $\langle n({\rm H}_2)\rangle$ can be replaced with $N({\rm H}_3^+)/L$ and $N({\rm H}_2)/L$, respectively (where $L$ is the cloud path length), thus putting equation (\ref{eqzeta2numden}) in terms of observables. As H$_3^+$ will form wherever there is an appreciable amount of H$_2$, it is reasonable to assume that the path length for both species is the same, such that
%\begin{equation}
%\zeta_2=k_{e}x_{e}n_{\rm H}\frac{N({\rm H}_3^+)}{N({\rm H}_2)}.
%\label{eqzeta2}
%\end{equation}
%Because the ratio $n({\rm H}_3^+)/n({\rm H}_2)$ is not expected to vary widely in models of diffuse molecular clouds \citep[e.g.,][]{neufeld2005}, this gives a representative value of the ionization rate throughout the entire cloud.

While equation (\ref{eqzeta2}) only considers one formation and destruction mechanism for H$_3^+$, comparison to a more complete chemical reaction network shows that it is a robust approximation given diffuse cloud conditions.  Assuming steady-state for H$_2^+$ where destruction by electron recombination and charge transfer to protons (reactions \ref{re_H2+_e} and \ref{re_H2+_H}) are considered, and steady-state for H$_3^+$ where destruction by proton transfer to CO and O (reactions \ref{re_H3+_CO_HCO+}, \ref{re_H3+_CO_HOC+}, and \ref{re_H3+_O}) is accounted for, gives the equations
\begin{equation}
\zeta_2n({\rm H}_2)=k_{\ref{re_H2+_e}}n_en({\rm H}_2^+)+k_{\ref{re_H2+_H}}n({\rm H})n({\rm H}_2^+)+k_{\ref{re_H2_H2+}}n({\rm H}_2)n({\rm H}_2^+),
\label{eq_H2+ss}
\end{equation}
\begin{equation}
k_{\ref{re_H2_H2+}}n({\rm H}_2)n({\rm H}_2^+)=k_en_en({\rm H}_3^+)+k_{\rm CO}n({\rm CO})n({\rm H}_3^+)+k_{\rm O}n({\rm O})n({\rm H}_3^+).
\label{eq_H3+ss_eCO}
\end{equation}
Solving for the ionization rate and making similar substitutions as before results in
\begin{equation}
\zeta_2=\frac{N({\rm H}_3^+)}{N({\rm H}_2)}n_{\rm H}\left[k_{e}x_e+k_{\rm CO}x({\rm CO})+k_{\rm O}x({\rm O})\right]
\left[1+\frac{2k_{\ref{re_H2+_e}}x_e}{k_{\ref{re_H2_H2+}}f_{{\rm H}_2}}+\frac{2k_{\ref{re_H2+_H}}}{k_{\ref{re_H2_H2+}}}\left(\frac{1}{f_{{\rm H}_2}}-1\right)\right],
\label{eq_fullzeta}
\end{equation}
where relevant rate coefficients are given in Table \ref{table_H3+ratecoeff}, $k_{\rm CO}=k_{\ref{re_H3+_CO_HCO+}}+k_{\ref{re_H3+_CO_HOC+}}$, and $f_{{\rm H}_2}\equiv2n({\rm H}_2)/n_{\rm H}$ is the fraction of hydrogen nuclei in molecular form.

\begin{sidewaysfigure}
\centering
\includegraphics[width=8.0in]{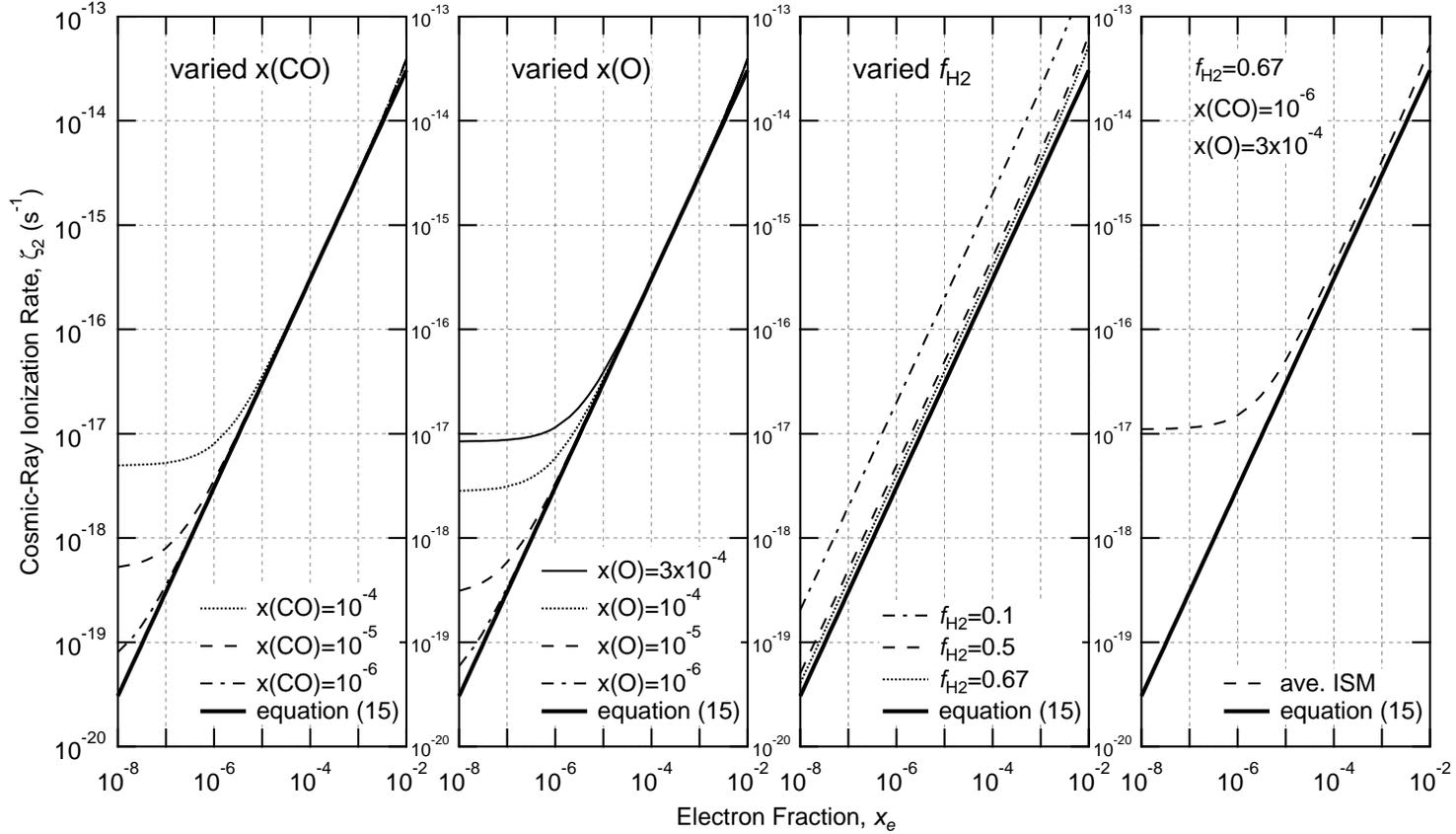}
\caption[Comparison between $\zeta_2$ derived from equation (\ref{eqzeta2}) versus equation (\ref{eq_fullzeta})]{\footnotesize Cosmic-ray ionization rate ($\zeta_2$) as a function of the electron fraction ($x_e$) for simplified chemistry (equation \ref{eqzeta2}) and more complete chemistry (equation \ref{eq_fullzeta}). In all cases $n_{\rm H}=200$ cm$^{-3}$, $N({\rm H}_3^+)/N({\rm H}_2)=10^{-7}$, and $T=70$ K.  The thick solid line shows the linear approximation given by equation (\ref{eqzeta2}).  The first panel shows the result of varying $x({\rm CO})$, with $x({\rm O})=10^{-8}$ and $f_{{\rm H}_2}=1$; the second panel shows the result of varying $x({\rm O})$, with $x({\rm CO})=10^{-8}$ and $f_{{\rm H}_2}=1$; the third panel shows the result of varying $f_{{\rm H}_2}$, with $x({\rm CO})=10^{-8}$ and $x({\rm O})=10^{-8}$; the last panel shows equation (\ref{eq_fullzeta}) for average diffuse ISM conditions ($f_{{\rm H}_2}=0.67$, $x({\rm CO})=10^{-6}$, $x({\rm O})=3\times10^{-4}$).  At $x_e=1.5\times10^{-4}$ equation (\ref{eq_fullzeta}) gives an ionization rate 1.33 times larger than equation (\ref{eqzeta2}), so if anything, we are slightly underestimating $\zeta_2$. \normalsize}
\label{fig_fullzeta}
\end{sidewaysfigure}

The cosmic-ray ionization rate as a function of electron fraction is plotted in Figure \ref{fig_fullzeta} for both equations (\ref{eqzeta2}) and (\ref{eq_fullzeta}).  In all four panels the thick solid line shows the linear relationship given by equation (\ref{eqzeta2}), while the various other curves show the ionization rate determined using equation (\ref{eq_fullzeta}) and different values of $x({\rm CO})$, $x({\rm O})$, and $f_{{\rm H}_2}$.  In the first panel $x({\rm CO})$ is set to $10^{-6}$, $10^{-5}$, and $10^{-4}$ with $x({\rm O})=10^{-8}$ and $f_{{\rm H}_2}=1$.  In the second panel $x({\rm O})$ is set to $10^{-6}$, $10^{-5}$, $10^{-4}$ and $3\times10^{-4}$ with $x({\rm CO})=10^{-8}$ and $f_{{\rm H}_2}=1$.  In the third panel $f_{{\rm H}_2}$ is set to 0.67, 0.5, and 0.1 with $x({\rm CO})=x({\rm O})=10^{-8}$.  

The first and second panels show that equation (\ref{eq_fullzeta}) only differs significantly from equation (\ref{eqzeta2}) when $x_e\lesssim 10^{-5}$, and for high fractional abundances of CO and O ($\gtrsim10^{-4}$).  These deviations occur when proton transfer to CO and O come to dominate over electron recombination as the primary destruction pathways for H$_3^+$.  The third panel shows that $f_{{\rm H}_2}$ affects the ionization rate for all values of $x_e$, but that rather low molecular hydrogen fractions are necessary to significantly alter the inferred value of $\zeta_2$.  With $f_{{\rm H}_2}=0.5$ (i.e., twice as many H atoms as H$_2$ molecules) the value output by equation (\ref{eq_fullzeta}) is about 1.6 times that from equation (\ref{eqzeta2}), while for $f_{{\rm H}_2}=0.67$ (i.e., equal number of H and H$_2$) the value output by equation (\ref{eq_fullzeta}) is about 1.3 times that from equation (\ref{eqzeta2}).  This deviation is caused by the larger relative abundance of atomic hydrogen destroying H$_2^+$ before it can form H$_3^+$, and is represented by the final term in equation (\ref{eq_fullzeta}).  The destruction of H$_2^+$ by electron recombination does not play a major role in the chemical network used here, and its influence can only be seen in the slight deviation between equations (\ref{eqzeta2}) and (\ref{eq_fullzeta}) when $x_e\sim10^{-2}$.  

Although some of the sight lines studied herein have molecular hydrogen fractions of $f_{{\rm H}_2}=0.2$ as determined from observations of H and H$_2$ \citep{savage1977,rachford2002,rachford2009}, it must be remembered that these are line-of-sight fractions, while equation (\ref{eq_fullzeta}) requires local fractions.  As it is expected that there will be atomic gas along a line of sight that is not associated with a cloud containing H$_2$, the line-of-sight value of $f_{{\rm H}_2}$ {\it always} underestimates the value of $f_{{\rm H}_2}$ within a diffuse molecular cloud.  It is generally assumed that the interior of a diffuse molecular cloud has conditions where $0.67<f_{{\rm H}_2}<1$ (i.e. somewhere between half and all of the hydrogen is in molecular form), such that reaction (\ref{re_H2+_H}) will not be very important.  

Using average diffuse molecular cloud conditions ($x({\rm O})=3\times10^{-4}$ \citet{cartledge2004,jensen2005}; $x({\rm CO})\sim10^{-6}$ \citet{sonnentrucker2007}; $f_{{\rm H}_2}\gtrsim0.67$) in equation (\ref{eq_fullzeta}) produces the dashed curve in the last panel of Figure \ref{fig_fullzeta}, and at $x_e=x({\rm C}^+)=1.5\times10^{-4}$ \citep{cardelli1996,sofia2004} the resulting ionization rate is only 1.33 times larger than that inferred from equation (\ref{eqzeta2}).  This demonstrates that equation (\ref{eqzeta2}) is a good approximation to equation (\ref{eq_fullzeta}) in the regions we study, and so we adopt equation (\ref{eqzeta2}) in computing $\zeta_2$.

With equation (\ref{eqzeta2}), the cosmic-ray ionization rate ($\zeta_2$) can be determined from the H$_3^+$ column density ($N({\rm H}_3^+)$), H$_2$ column density ($N({\rm H}_2)$), total hydrogen density ($n_{\rm H}$), electron fraction ($x_e$), and H$_3^+$-electron recombination rate coefficient ($k_e$).  The electron fraction can be approximated by the fractional abundance of C$^{+}$ assuming that nearly all electrons are the result of singly photoionized carbon.  In sight lines where C$^+$ has been observed, the measured value of $x({\rm C}^+)$ is used for $x_e$.  For all other sight lines, the average fractional abundance measured in various diffuse clouds, $x({\rm C}^+)\approx1.5\times10^{-4}$ \citep{cardelli1996,sofia2004}, is adopted for $x_e$.  Uncertainties in $x_e$ are assumed to be $\pm20$\%, i.e., $\pm3\times10^{-5}$.  The H$_3^+$-electron recombination rate coefficient has now been measured in multiple laboratory experiments \citep[e.g.,][]{mccall2004,kreckel2005,kreckel2010} with consistent results and is presented in Table \ref{table_H3+ratecoeff}.  When available the spin temperature of H$_2$, $T_{01}$, and its uncertainty are used in calculating $k_e$; otherwise, an average value of $70\pm10$~K is adopted.  The total hydrogen number density is difficult to determine, but various studies have estimated $n_{\rm H}$ using a rotational excitation analysis of observed C$_2$ lines \citep{sonnentrucker2007}, an analysis of H and the $J=4$ level of H$_2$ \citep{jura1975}, or a thermal pressure analysis of fine structure lines of C~\textsc{i} \citep{jenkins1983}.  Number densities from these studies are presented in Table \ref{table_zeta2} when available.  For sight lines without estimated densities, the rough average value of $n_{\rm H}=200$ cm$^{-3}$ is adopted.  In all cases, uncertainties in $n_{\rm H}$ are assumed to be $\pm50\%$ of the reported values.  Absorption lines from electronic transitions of H$_2$ have been observed in the UV along many of the sight lines in this study \citep[from which the aforementioned values of $T_{01}$ are derived;][]{savage1977,rachford2002,rachford2009}.  In sight lines where H$_2$ has not been observed, two other methods were used to estimate $N({\rm H}_2)$.  The preferred method uses column densities of CH determined from observations in combination with the relation $N({\rm CH})/N({\rm H}_2)=3.5^{+2.1}_{-1.4}\times10^{-8}$ from \citet{sheffer2008}.  In sight lines where neither H$_2$ nor CH has been observed, $N({\rm H}_2)$ is estimated from the color excess, $E(B-V)$, using the relation $N_{\rm H}\approx E(B-V)5.8\times10^{21}$ cm$^{-2}$ mag$^{-1}$ from \citet{bohlin1978}, and assuming $f_{{\rm H}_2}\approx2N({\rm H}_2)/N_{\rm H}=0.67$.  Molecular hydrogen column densities determined both from observations and estimates are presented in Table \ref{table_zeta2}.

While Table \ref{table_lineparam} gives individual column densities for the lowest lying {\it ortho} and {\it para} levels of H$_3^+$, values of $N({\rm H}_3^+)$ in Table \ref{table_zeta2} are equal to the sum of the column densities in the $(1,1)$ and $(1,0)$ states (i.e., $N({\rm H}_3^+)=N(1,1)+N(1,0)$).  In cases where observations of both the $R(1,1)^u$ and $R(1,1)^l$ lines produce independent values of $N(1,1)$, a variance weighted average,
\begin{equation}
N=\sum_{i=1}^{n} (N_i/\sigma_i^2)/\sum_{i=1}^{n} (1/\sigma_i^2),
\label{eq_var_weight_column}
\end{equation}
is used to determine $N(1,1)$. Because there are only 2 independent measurements of $N(1,1)$, the uncertainty of the weighted average, $S[N(1,1)]$, is determined using an unbiased estimator of a weighted population variance for small samples.  This is given by
\begin{equation}
S^2=\frac{V_1}{V_1^2-V_2}\sum_{i=1}^{n}w_i(N_i-\mu^*)^2,
\label{eq_unbiased uncertainty}
\end{equation}
where
%\begin{eqnarray}
%w_i=\frac{1}{\sigma_i^2}, \\
%V_1=\sum_{i=1}^{n}w_i, \\
%V_2=\sum_{i=1}^{n}w_i^2,
%\label{eq_unbiased_parts}
%\end{eqnarray}
\begin{equation*}
w_i=\frac{1}{\sigma_i^2},~~V_1=\sum_{i=1}^{n}w_i,~~V_2=\sum_{i=1}^{n}w_i^2,
\label{eq_unbiased_parts}
\end{equation*}
and $\mu^*$ is the weighted average determined from equation (\ref{eq_var_weight_column}).  The uncertainty in the total H$_3^+$ column density is then computed as usual by adding $\sigma[N(1,1)]$ and $\sigma[N(1,0)]$ in quadrature.  Upper limits to the H$_3^+$ column density should be taken as $3\sigma[N({\rm H}_3^+)]$.

Using equation (\ref{eqzeta2}), and taking the values described above and in Table \ref{table_zeta2}, cosmic-ray ionization rates are inferred for all diffuse molecular clouds where H$_3^+$ observations have been made.  These values of $\zeta_2$ are presented in column 10 of Table \ref{table_zeta2}, with uncertainties in column 11.  As for the H$_3^+$ column densities, upper limits to the cosmic-ray ionization rate should be taken as $3\sigma(\zeta_2)$.

For one sight line however, that toward NGC 2024 IRS 1, a different analysis is used because recent observations suggest that the interstellar material is more likely dense than diffuse (T.~Snow 2011, private communication).  In this case, values appropriate for dense clouds ($x_e=10^{-7}$, $f_{{\rm H}_2}=1$) are adopted, effectively simplifying equation (\ref{eq_fullzeta}) to
\begin{equation}
\zeta_2=\frac{N({\rm H}_3^+)}{N({\rm H}_2)}n_{\rm H}\left[k_{\rm CO}x({\rm CO})+k_{\rm O}x({\rm O})\right].
\label{eq_zeta_dense}
\end{equation}
The density and temperature are also set to average dense cloud values ($n_{\rm H}=10^4$ cm$^{-3}$, $T=30$ K).  The large CO column density ($N({\rm CO})=1.26\times10^{18}$ cm$^{-2}$; T.~Snow 2011, private communication) results in $x({\rm CO})=1.28\times10^{-4}$, demonstrating that most of the carbon is in molecular form, and thus validating the low value of $x_e$ assumed above.  Oxygen abundances are typically about two times carbon abundances \citep{lodders2003}, such that half of all O is expected to be in the form of CO.  We assume the remainder to be in atomic form, and use $x({\rm O})=x({\rm CO})$ in equation (\ref{eq_zeta_dense}).  The resulting upper limit on the cosmic-ray ionization rate toward NGC 2024 IRS 1 is $\zeta_2<4.2\times10^{-17}$ s$^{-1}$.  This lower value is expected in dense cloud conditions, as will be discussed below, and is excluded from our analysis of the distribution of ionization rates in diffuse clouds.

\subsection{The Distribution of Cosmic-Ray Ionization Rates in Diffuse Clouds}

\begin{figure}
\centering
\includegraphics[width=6.5in]{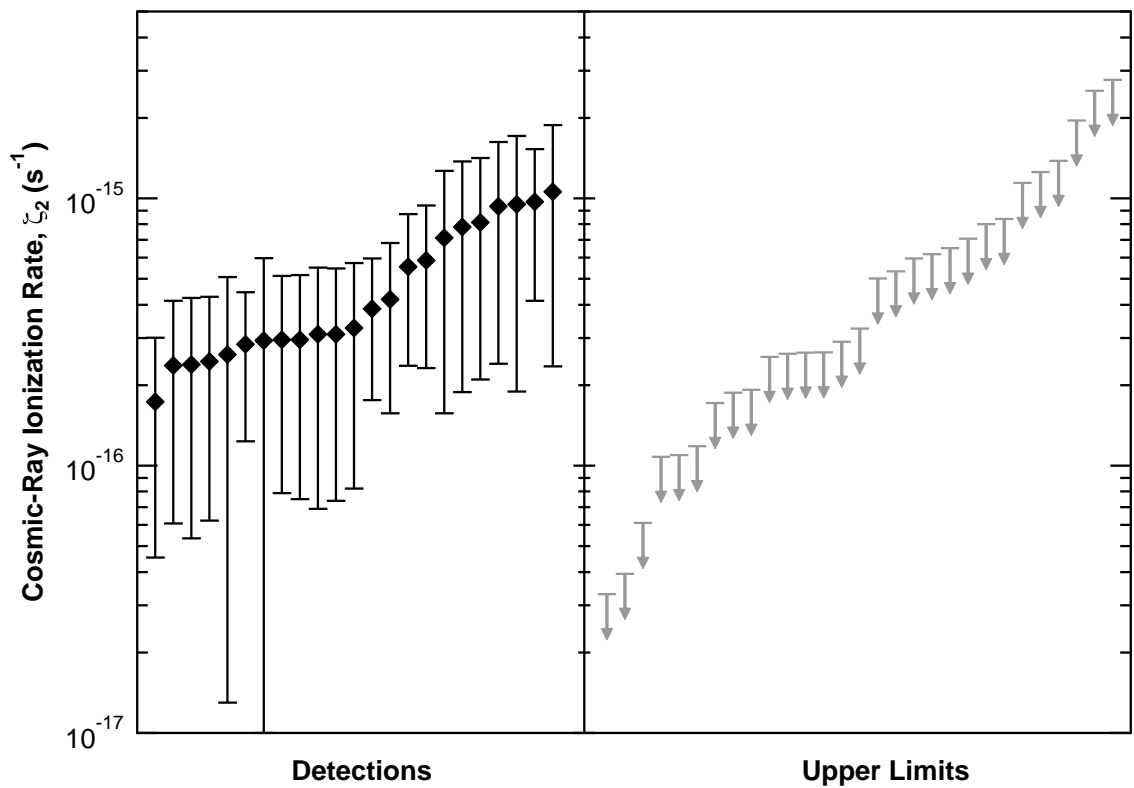}
\caption[Cosmic-ray ionization rates and upper limits]{Cosmic-ray ionization rates and upper limits from the sample of diffuse cloud sight lines.  The left panel shows ionization rates inferred from detections of H$_3^+$ with $1\sigma$ uncertainties.  Points are spread out horizontally for clarity.  The right panel shows $3\sigma$ upper limits to the ionization rate inferred for sight lines where H$_3^+$ is not detected.  Some of the $3\sigma$ upper limits suggest ionization rates lower than those inferred in sight lines with H$_3^+$ detections.  This likely points to variations in the cosmic-ray ionization rate between sight lines.
}
\label{fig_zeta2_withmean}
\end{figure}

With ionization rates and upper limits inferred for 50 diffuse cloud sight lines, it is interesting to study the distribution of our entire sample.  Cosmic-ray ionization rates with $1\sigma$ uncertainties and $3\sigma$ upper limits are plotted in Figure \ref{fig_zeta2_withmean}.  In order to find the probability density function for the cosmic-ray ionization rate in our sample, all of these data points are combined.  Each sight line with an H$_3^+$ detection is treated as a gaussian function with mean and standard deviation equal to the ionization rate and uncertainty reported in columns 10 and 11 of Table \ref{table_zeta2}.  Each sight line with a non-detection of H$_3^+$ is treated as a gaussian function with mean equal to zero and standard deviation equal to the uncertainty in column 11 of Table \ref{table_zeta2}.  The cosmic-ray ionization rate cannot be negative, and it is reasonable to assume that in diffuse clouds there is some ``floor'' value below which $\zeta_2$ does not fall caused by the continuous diffusion of particles throughout the Galactic disk \citep{webber1998}.  We set this floor value to $10^{-17}$~s$^{-1}$ --- a rough average of ionization rates computed from proposed local interstellar proton spectra \citep[e.g.,][]{spitzer1968,webber1998} --- and truncate all of the gaussian functions at this floor.\footnote{Setting the floor ionization rate to zero does not significantly change our results.}  Each function is then re-normalized so that the area under the curve above the floor ionization rate is equal to 1.  The truncated gaussians are added together, and the final distribution is again normalized to have an area equal to 1.  

\begin{figure}
\centering
\includegraphics[width=4.0in]{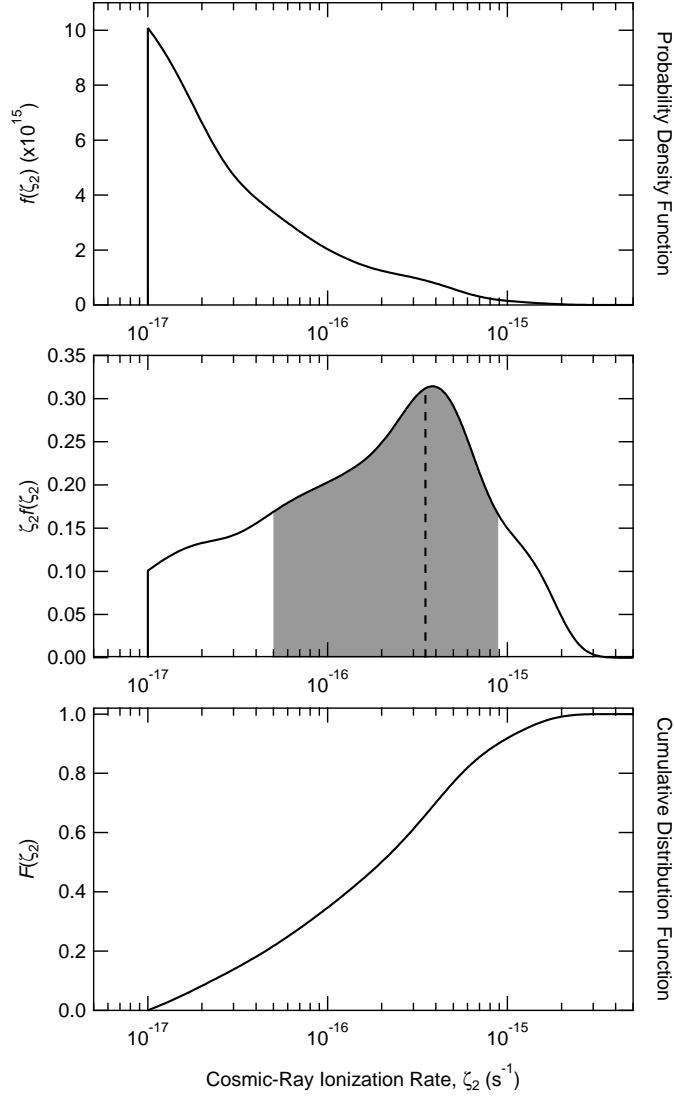}
\caption{Upper panel: Probability density function, $f(\zeta_2)$, found by combining inferred ionization rates and upper limits.  This is defined such that the probability of the cosmic-ray ionization rate being in any particular range is given by $P(a\leq\zeta_2\leq b)=\int_{a}^{b}f(\zeta_2)d\zeta_2$.  Center panel: Probability density function multiplied by the ionization rate, $\zeta_2f(\zeta_2)$.  This allows one to see ``by eye" what portion of $f(\zeta_2)$ carries the largest probability given the logarithmic $x$-axis.  The vertical dashed line shows the mean value of $\zeta_2=3.5\times10^{-16}$~s$^{-1}$, and the shaded region is the smallest range of ionization rates in log space that contains 68.3\% of the area under the probability density function.  Lower panel: Cumulative distribution function, $F(\zeta_2)$, showing the probability that $\zeta_2$ is below any given value.}
\label{fig_stats}
\end{figure}

The resulting probability density function, $f(\zeta_2)$, is plotted in the top panel of Figure \ref{fig_stats}.  Because of the logarithmic $x$-axis, it is difficult to tell ``by eye'' which ionization rates are most probable.  However, this is easily seen by scaling $f(\zeta_2)$ by $\zeta_2$, as shown in the middle panel of Figure \ref{fig_stats}.  By integrating $\zeta_2f(\zeta_2)$, we find the mean value of the cosmic-ray ionization rate to be $3.5\times10^{-16}$~s$^{-1}$, marked by the vertical dashed line.  The minimum range of ionization rates (in log space) that contains 68.3\% (gaussian 1$\sigma$ equivalent) of the area under $f(\zeta_2)$ is bounded by $5.0\times10^{-17}$~s$^{-1}$ on the left and $8.8\times10^{-16}$~s$^{-1}$ on the right, and is shown by the shaded region.  The bottom panel of Figure \ref{fig_stats} shows the cumulative distribution function, $F(\zeta_2)$, which gives the probability that the ionization rate is below a particular value.

%, resulting in the probability density, $f(\zeta_2)$, and cumulative distribution, $F(\zeta_2)$, functions plotted in the top and bottom panels of Figure \ref{fig_stats}, respectively.  Because of the logarithmic $x$-axis, it is much easier to see where the ``weight'' of the distribution lies by looking at $\zeta_2f(\zeta_2)$, as shown in the middle panel.  By integrating $\zeta_2f(\zeta_2)$, we find the mean value of the cosmic-ray ionization rate to be $3.5\times10^{-16}$~s$^{-1}$.  The minimum range of ionization rates (in log space) that contains 68.3\% (gaussian 1$\sigma$ equivalent) of the area under $f(\zeta_2)$ is bounded by $5.0\times10^{-17}$~s$^{-1}$ on the left and $8.8\times10^{-16}$~s$^{-1}$ on the right, and is shown by the shaded region in the middle panel of Figure \ref{fig_stats}.

\section{CORRELATIONS BETWEEN LINE-OF-SIGHT PROPERTIES}

\subsection{The {\rm H}$_3^+$ Column Density}

\begin{figure}
\centering
\includegraphics[width=6.5in]{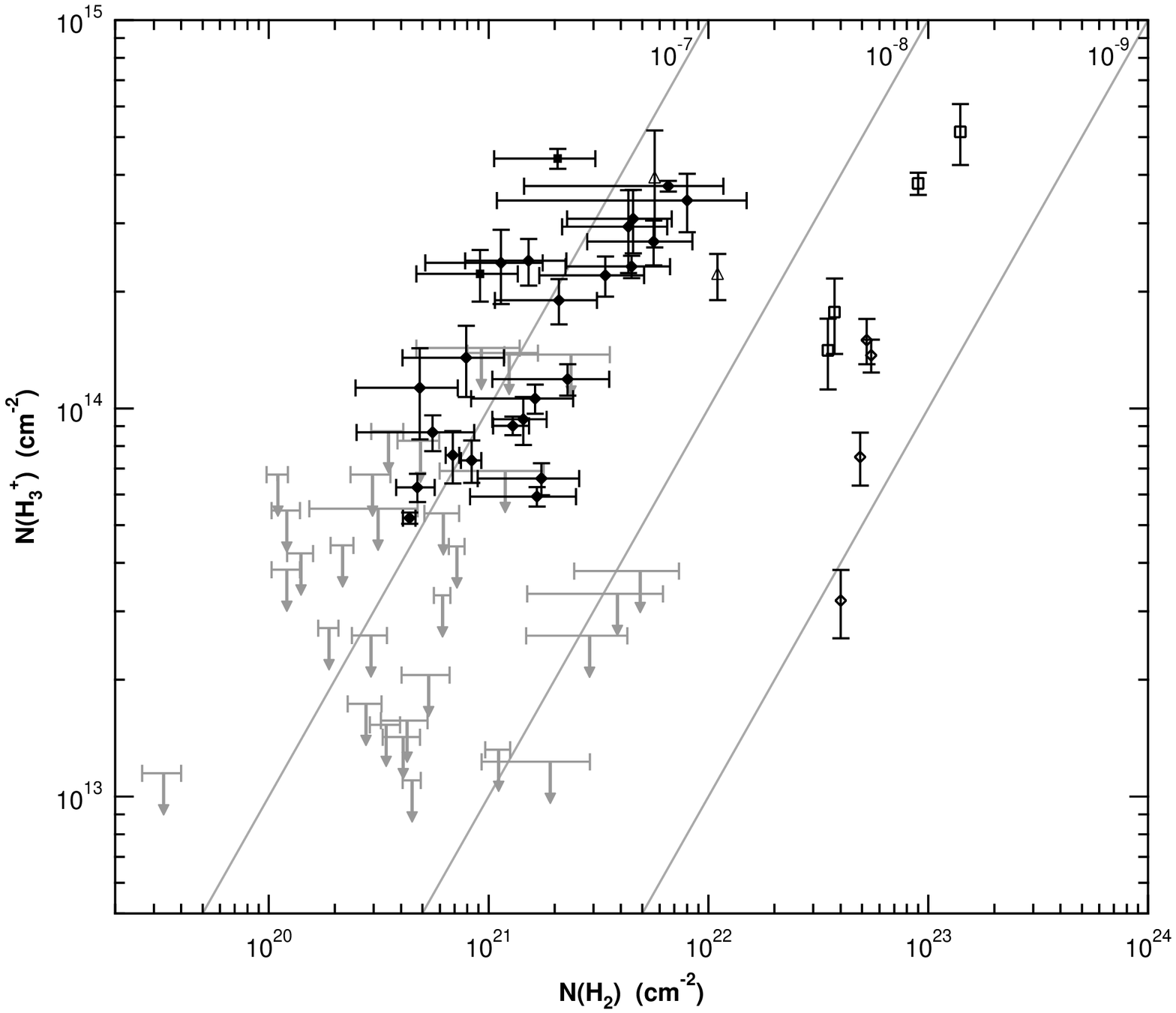}
\caption[$N({\rm H}_3^+)$ versus $N({\rm H}_2)$]{$N({\rm H}_3^+)$ versus $N({\rm H}_2)$ for diffuse clouds studied herein and various dense clouds reported in the literature.  Symbols are as follows: filled diamonds - this study (detections); filled squares - \citet{indriolo2010b}; grey arrows - this study ($3\sigma$ upper limits on $N({\rm H}_3^+)$); open squares - \citet{mccall1999}; open diamonds - \citet{kulesa2002}; open triangles - \citet{brittain2004} and \citet{gibb2010}.  The diagonal lines show constant values of $N({\rm H}_3^+)/N({\rm H}_2)$ and are labeled accordingly.  Diffuse clouds cluster about $N({\rm H}_3^+)/N({\rm H}_2)=10^{-7}$, while dense clouds fall in the range between $10^{-8}$ and $10^{-9}$.  For diffuse cloud sight lines $N({\rm H}_2)$ is determined from UV H$_2$ observations, estimated from $N({\rm CH})$, or estimated from $E(B-V)$, and values used here are presented in Table \ref{table_zeta2}.  For dense cloud sight lines $N({\rm H}_2)$ is determined from IR H$_2$ observations, estimated from $A_V$, or estimated from $N({\rm CO})$.
}
\label{fig_H3+_vs_H2}
\end{figure}

Given that the H$_3^+$ formation process starts with H$_2$, one might expect that abundances of the 2 species should be correlated.  Figure \ref{fig_H3+_vs_H2} shows $N({\rm H}_3^+)$ versus $N({\rm H}_2)$ for the entire sample of sight lines studied herein.  Also included are H$_3^+$ column densities in sight lines near the supernova remnant IC 443 \citep{indriolo2010b}, as well as in dense clouds \citep{mccall1999,kulesa2002,brittain2004,gibb2010}.  It is clear that for diffuse cloud sight lines where H$_3^+$ has been detected, relative abundances tend to cluster about $10^{-7}$.  A re-arrangement of equation (\ref{eqzeta2}),
\begin{equation*}
\frac{N({\rm H}_3^+)}{N({\rm H}_2)}=\frac{\zeta_2}{k_{e}x_{e}n_{\rm H}},
\label{eq_H3+overH2}
\end{equation*}
suggests that scatter about a central value of $N({\rm H}_3^+)/N({\rm H}_2)$ is likely due to variations in $\zeta_2$ and electron density ($x_en_{\rm H}$) between sight lines.
 
The much lower value of $N({\rm H}_3^+)/N({\rm H}_2)$ in dense clouds ($10^{-8}$--$10^{-9}$) is the result of a higher density, different destruction partner, and lower ionization rate.  Replacing electrons with CO as the dominant destruction partner leaves $N({\rm H}_3^+)/N({\rm H}_2)$ inversely proportional to $k_{\rm CO}x({\rm CO})n_{\rm H}$.  The rate coefficient for destruction of H$_3^+$ by CO is about 2 orders of magnitude slower than by electrons (see Table \ref{table_H3+ratecoeff}), the density of dense clouds is about 2--3 orders of magnitude larger than diffuse clouds \citep{snow2006}, and the fractional abundance of electrons in diffuse clouds should be about equal to that of CO in dense clouds.  This means that $\zeta_2$ must be about 0--2 orders of magnitude lower in dense clouds than in diffuse clouds to produce the observed results, and several studies of ionized species in dense clouds support such a trend \citep[e.g., ][]{vandertak2000,kulesa2002,hezareh2008}.

\subsection{The Cosmic-Ray Ionization Rate and Location}

\begin{sidewaysfigure}
\centering
\includegraphics[width=8.9in]{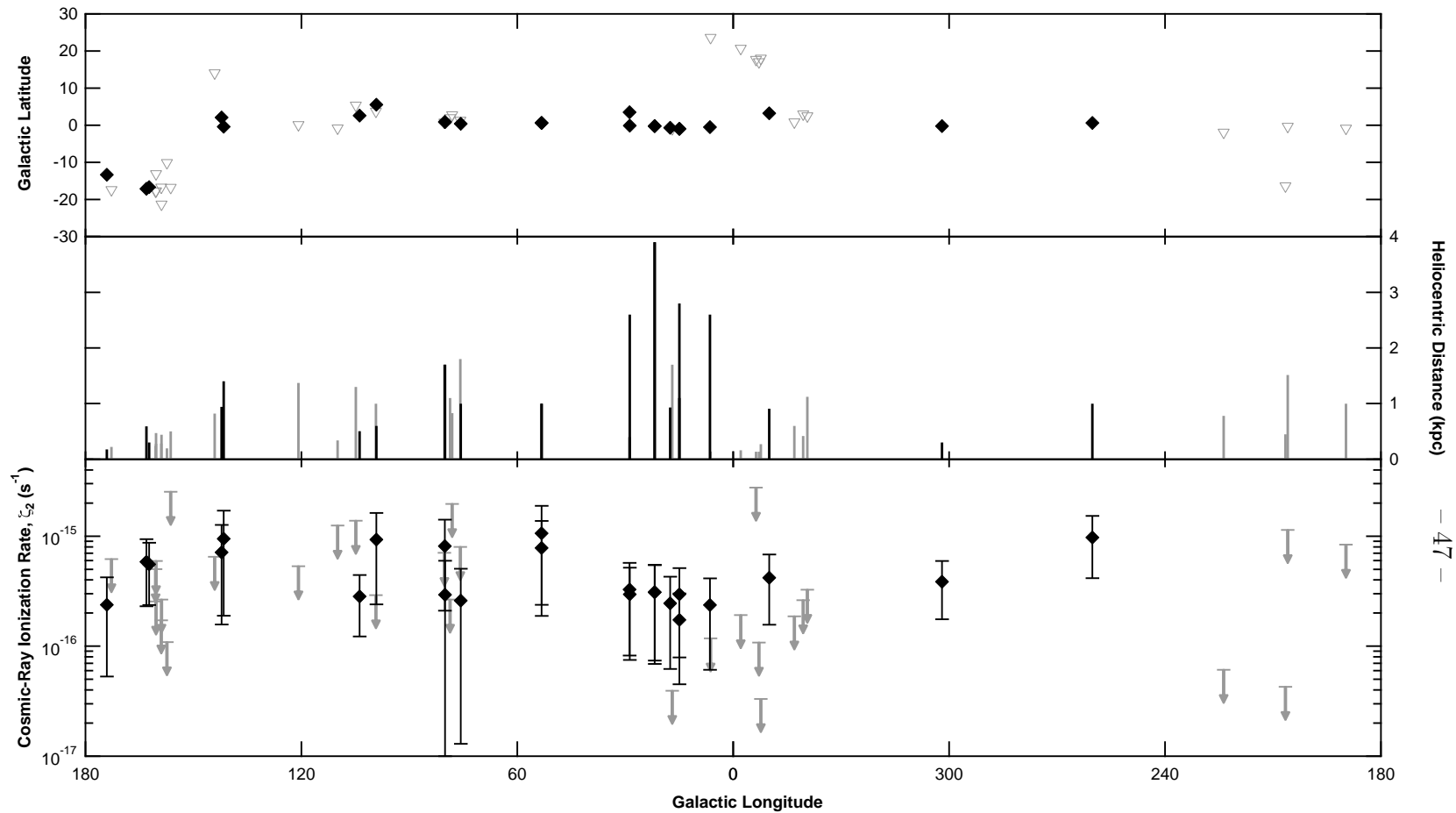}
\caption[Cosmic-ray ionization rate with respect to Galactic coordinates and heliocentric distance]{{\bf Top panel:} Map showing the locations of targeted sight lines in Galactic coordinates.  Black diamonds mark sight lines where H$_3^+$ absorption is observed, and open grey triangles mark sight lines where H$_3^+$ is not detected. {\bf Middle panel:} Heliocentric distance of background sources with respect to Galactic longitude.  Black lines mark sight lines where H$_3^+$ absorption is observed, and grey lines mark sight lines where H$_3^+$ is not detected.  {\bf Bottom panel:} The cosmic-ray ionization rate as a function of Galactic longitude.  Black diamonds are inferred cosmic-ray ionization rates with $1\sigma$ uncertainties.  Grey arrows are $3\sigma$ upper limits.}
\label{fig_zeta2_glat_dist_vs_glong}
\end{sidewaysfigure}

Beyond studying the cosmic-ray ionization rate in and of itself, it is insightful to search for correlations between $\zeta_2$ and various other sight line parameters.  One of the most fundamental relationships to study is that between $\zeta_2$ and location within the Galaxy.  The cosmic-ray ionization rate is plotted against Galactic longitude in the bottom panel of Figure \ref{fig_zeta2_glat_dist_vs_glong}.  The top panel is a map in Galactic coordinates and the middle panel gives the distance to background sources such that the reader can easily trace the ionization rates in the bottom panel to an actual on-sky position.  There are no apparent gradients in Figure \ref{fig_zeta2_glat_dist_vs_glong}, and comparisons of $\zeta_2$ with both heliocentric distance and Galactocentric radius of background sources do not show any trends, thus suggesting that the mechanism responsible for variations in the cosmic-ray ionization rate does not operate on large scales.  To study variations of $\zeta_2$ on small spatial scales, we investigate some of the closest sight line pairings from our data.

\subsubsection{HD 168607 \& HD 168625}

The sight lines with the smallest angular separation in our survey are HD~168607 and HD~168625, separated by only 1\farcm1.  The H$_3^+$ spectra for both targets are very similar (see Figure \ref{fig_ukirt_det2}), suggesting that the absorption arises from a common cloud that must be in front of the closer target (HD 168607) at 1100~pc.  At this distance, 1\farcm1 corresponds to an on-sky separation of 0.35~pc.  Ionization rates for these two sight lines are consistent with each other within uncertainties, demonstrating uniformity of $\zeta_2$ on the smallest spatial scales probed by our observations.

\subsubsection{Cyg OB2 Association}

Three of our target sight lines use stars in the Cyg OB2 association as background sources: Cyg OB2 5, Cyg OB2 12, and Cyg OB2 8A.  Taking the OB association to be at 1.7~kpc \citep{torresdodgen1991}, the on-sky distance between No. 5 and No. 12 is 2.6~pc, between No. 5 and No. 8A is 4.9~pc, and between No. 12 and No. 8A is 3.8~pc.  Spectra toward No. 5 and No. 12 are somewhat similar, with components at about 6~km~s$^{-1}$ and 12~km~s$^{-1}$ LSR in a few molecular species \citep{mccall2002}.  While the ionization rates in these two sight lines differ by about a factor of 3, the inability to meaningfully separate the H$_3^+$ absorption features into the different velocity components makes this analysis very uncertain, and $\zeta_2$ for Cyg OB2 5 and Cyg OB2 12 have overlapping 1$\sigma$ uncertainties.  Cyg OB2 8A, although not far removed from the other two sight lines, does not show H$_3^+$ absorption (see Figure \ref{fig_keck2}), and may only contain the 6~km~s$^{-1}$ component \citep{snow2010}.  Still, the $3\sigma$ upper limit on $\zeta_2$ toward No. 8A is consistent with the ionization rates inferred for the other two sight lines, and nothing conclusive can be drawn from this region.  

A complementary method of tracing the cosmic-ray flux relies on gamma-ray emission from $\pi^0$ decay, and observations of the Cygnus X region by {\it Fermi} LAT are suggestive of a large flux of recently accelerated hadronic cosmic rays \citep{ackermann2011} near the Cyg OB2 association.  This is one of a few select regions where both gamma-rays and H$_3^+$ have been observed, and presents the opportunity for a rather unique study based on both cosmic ray tracers.  However, this analysis is beyond the scope of the current paper, and such a study would best be served by an expanded survey of H$_3^+$ that probes a larger extent of the gas traced by the gamma-ray emission presented in \citet{ackermann2011}.

\subsubsection{Per OB2 Association}

The targets $\zeta$~Per, X~Per, and $o$~Per are all in the Perseus OB2 association \citep[although the latter two may only be in the association by projection;][]{dezeeuw1999}.  X~Per and $\zeta$~Per show very similar spectra (see Figure \ref{fig_ukirt_det1}) with H$_3^+$ absorption lines at 7--8~km~s$^{-1}$ LSR, and the ionization rate inferred for each sight line is nearly identical.  At a distance of 301~pc, the angular separation of 0\fdg88 corresponds to 4.6~pc in the plane of the sky.  Among our targeted sight lines $o$~Per is the next closest to these two, and, assuming a distance of 283~pc, is 10.5~pc away from $\zeta$~Per and 13.2~pc away from X~Per.  H$_3^+$ is not detected toward $o$~Per, and the $3\sigma$ upper limit inferred for $\zeta_2$ in this sight line is about equal to the lower $1\sigma$ bounds on the ionization rates toward X~Per and $\zeta$~Per.  While $o$~Per is about 20~pc closer than $\zeta$~Per, the similar $E(B-V)$ values for both sight lines make it unlikely that we do not see H$_3^+$ simply because it is behind the target.  Although not definitive, these results strongly suggest variations in the cosmic-ray ionization rate on a size scale of order 10~pc.

\subsubsection{Ophiuchus-Scorpius Region}

Many of our target sight lines probe the nearby Ophiuchus-Scorpius region, and some of the most closely spaced include $o$~Sco, $\rho$~Oph~D, HD~147889, and $\chi$~Oph.  The first three are all separated by 2--3~pc (assuming a distance of 136~pc), with $\chi$~Oph about 12--14~pc away.  This is one of a few regions where H$_3^+$ is not detected in any of the target sight lines, and aside from $\rho$~Oph~D where the low continuum S/N level ($\sim175$) of the spectrum limits our analysis, the $3\sigma$ upper limits on $\zeta_2$ are in the range $3$--$19\times10^{-17}$~s$^{-1}$.  These are some of the smaller inferred upper limits on $\zeta_2$, and likely pertain to the material closest to the sun probed in our survey as background sources are only 100--200~pc away.  It is interesting then, that these upper limits are consistent with the ionization rate of a few times $10^{-17}$~s$^{-1}$ predicted by the local interstellar cosmic-ray spectrum \citep{webber1998}.

\subsection{The Cosmic-Ray Ionization Rate and Hydrogen Column Density}

\begin{sidewaysfigure}
\centering
\includegraphics[width=8.0in]{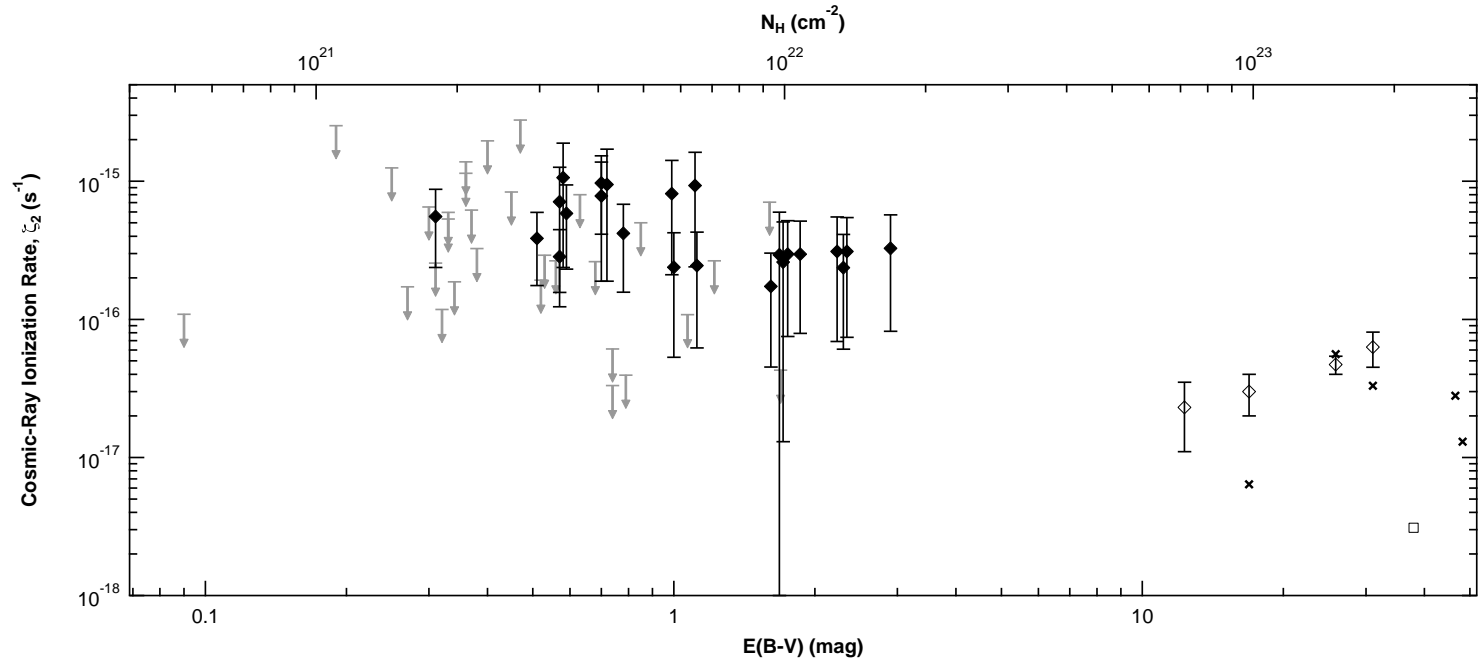}
\caption[Cosmic-ray ionization rate versus total column density]{The cosmic-ray ionization rate as a function of cloud column density, $N_{\rm H}$, and equivalent color excess, $E(B-V)$.  For sight lines with multiple distinct velocity components, the total line-of-sight $N_{\rm H}$ has been divided by the number of cloud components.  Filled diamonds are cosmic-ray ionization rates with $1\sigma$ uncertainties inferred for diffuse molecular clouds in this study.  Grey arrows are $3\sigma$ upper limits from this study.  Ionization rates inferred in dense clouds are as follows: open diamonds - H$_3^+$ observations of NGC 2024 IRS 2, AFGL 490, AFGL 2591, and AFGL 2136 by \citet{kulesa2002}; crosses - HCO$^+$ observations of AFGL 2136, AFGL 2591, AFGL 490, W33 A, and W3 IRS 5 by \citet{vandertak2000}; open square - HCNH$^+$ observation of DR 21(OH) by \citet{hezareh2008}.
%Open diamonds are ionization rates inferred from H$_3^+$ observations for dense clouds by \citet{kulesa2002}.  Sight lines and ionization rates from \citet{kulesa2002} include: NGC 2024 IRS 2 - $2.3\pm1.7\times10^{-17}$ s$^{-1}$; AFGL 490 - $3\pm1\times10^{-17}$ s$^{-1}$; AFGL 2591 - $4.7\pm0.7\times10^{-17}$ s$^{-1}$; AFGL 2136 - $6.3\pm1.8\times10^{-17}$ s$^{-1}$. The crosses are ionization rates inferred from H$^{13}$CO$^+$ observations in dense clouds by \citet{vandertak2000}. Sight lines and ionization rates from \citet{vandertak2000} include: AFGL 2136 - $3.3\times10^{-17}$ s$^{-1}$; AFGL 2591 - $5.6\times10^{-17}$ s$^{-1}$; AFGL 490 - $0.64\times10^{-17}$ s$^{-1}$; W33 A - $1.3\times10^{-17}$ s$^{-1}$; W3 IRS 5 - $2.8\times10^{-17}$ s$^{-1}$;  The open square is the ionization rate inferred from HCNH$^+$ observations of DR 21(OH) by \citet{hezareh2008}.
Various models \citep[e.g.,][]{padovani2009} predict that $\zeta_2$ should decrease with increased $N_{\rm H}$ as low-energy cosmic rays lose all of their energy in the outer regions of a cloud.  While no strong correlation is apparent from the diffuse cloud data alone, the addition of the dense cloud ionization rates from sight lines with much higher $N_{\rm H}$ does suggest such a trend.
}
\label{fig_zeta2_vs_EBV}
\end{sidewaysfigure}

Given that there is no evidence for large-scale gradients in the cosmic-ray ionization rate, it is possible that properties of the observed clouds themselves are responsible for variations in $\zeta_2$, and will show some correlation with the cosmic-ray ionization rate.  A parameter with which $\zeta_2$ has been predicted to vary is the hydrogen column density, $N_{\rm H}$ \citep{padovani2009}.  This is because the energy spectrum of cosmic-rays is expected to change with depth into a cloud.  Lower-energy particles --- those most efficient at ionization --- will lose all of their energy to ionization interactions in the outer regions of a cloud, leaving only higher-energy particles to ionize the cloud interior.  As such, the ionization rate in a cloud interior should be lower than in a cloud exterior.  Because lower-energy particles can operate through a larger portion of clouds with lower column densities \citep{cravens1978}, it is expected that the inferred $\zeta_2$ will decrease with increasing $N_{\rm H}$.  A plot of $\zeta_2$ versus $N_{\rm H}$ (and equivalent $E(B-V)$) is shown in Figure \ref{fig_zeta2_vs_EBV}.  In sight lines with multiple distinct velocity components, the total line-of-sight $N_{\rm H}$ has been divided by the number of components to better approximate the size of each individual cloud.  Included in Figure \ref{fig_zeta2_vs_EBV} are data from 4 dense cloud sight lines observed in H$_3^+$ by \citet{kulesa2002}, 5 dense cloud sight lines observed in H$^{13}$CO$^+$ by \citet{vandertak2000}, and 1 dense cloud sight line observed in HCNH$^+$ by \citet{hezareh2008} for the purpose of extending the relationship to much higher $N_{\rm H}$.  There does not appear to be a correlation between $\zeta_2$ and $N_{\rm H}$ when only the diffuse molecular cloud sight lines are considered. However, when the dense cloud ionization rates (most of which are of order a few times $10^{-17}$ s$^{-1}$) are included, it seems that $\zeta_2$ does decrease in sight lines with higher column densities.

\section{DISCUSSION}

Given that $\zeta_2$ does {\it not} decrease with increased $N_{\rm H}$ among diffuse cloud sight lines, but does when switching from the diffuse to dense cloud regime, the following conclusions can be drawn.  First, the cosmic rays that are primarily responsible for ionization in diffuse clouds must be able to propagate entirely through such clouds.  This could easily be done by 10 MeV protons, which have a range of $Rn\approx2\times10^{22}$ cm$^{-2}$ \citep{cravens1978}, about equal to the largest values of $N_{\rm H}$ in diffuse clouds studied here.  However, a cloud with a large line-of-sight column density does not necessarily have a large column density in the plane of the sky.  Even lower-energy particles then (e.g., 2 MeV protons with $Rn\approx10^{21}$ cm$^{-2}$) could potentially cause ionization through the entire extent of an observed cloud.

Second, the cosmic rays that are responsible for most of the ionization in diffuse clouds must not be able to penetrate to the interiors of dense clouds.  Again, something like 2--10 MeV protons fit this picture well.  Most of the sight lines where H$_3^+$ is observed in dense clouds use embedded objects as background sources.  Regions surrounding these sources have been mapped in various molecular species (via emission at radio wavelengths), implying large column densities in the plane of the sky in addition to the large column densities observed along the line of sight.  This increases the likelihood that much of the material being probed is deep within the observed cloud.

Taken together, the higher ionization rates inferred in diffuse clouds and lower ionization rates inferred in dense clouds suggest that differences in $\zeta_2$ can be explained by the inability of low-energy cosmic rays to penetrate large columns of material.  However, under the assumption that the cosmic-ray spectrum is uniform throughout the Galactic disk \citep{webber1998}, this does not adequately explain differences in the cosmic-ray ionization rate amongst diffuse clouds.  Instead, it would seem that the cosmic-ray spectrum --- at least for particles in the energy range most efficient at ionization --- must vary in space.

Despite the previously held assumption of a uniform cosmic-ray spectrum, it should not be surprising that the flux of low-energy particles varies across the Galaxy.  Given a hydrogen density of 1 cm$^{-3}$, a 2 MeV proton will only travel about 320 pc (not necessarily in a straight line) before losing all of its energy, meaning that any point that is a few hundred parsecs away from a site of particle acceleration will not experience the same flux of 2 MeV protons as a point that is much closer to an acceleration site.  For low-energy particles to even enter a diffuse cloud then, the cloud must be relatively close to a site of particle acceleration.

Regardless of where most low-energy cosmic rays are accelerated, it is possible that differences in $\zeta_2$ among diffuse cloud sight lines can be attributed to the distance between a cloud and acceleration site.  Sight lines that probe material in close proximity to an acceleration site should show high ionization rates, while those that probe material farther away should show lower ionization rates.  Unfortunately, defining acceleration sites and actually computing physical distances between those and the sample of observed clouds is difficult at best.  

There are a few special cases, however, where the ionization rate can be inferred in molecular clouds known to be in close proximity to a site of cosmic-ray acceleration such as a supernova remnant.  Shocked gas \citep[e.g.,][]{huang1986,dickman1992} and OH (1720 MHz) masers \citep{claussen1997,hewitt2006} observed toward the supernova remnant IC 443 demonstrate that the expanding shell is physically interacting with a foreground cloud, and gamma-ray observations of the remnant \citep{albert2007,acciari2009,abdo2010ic443,tavani2010} show a signature indicative of $\pi^0$ decay that suggests the cloud is experiencing a large flux of energetic protons.  In \citet{indriolo2010b} we presented observations of H$_3^+$ in 6 sight lines that pass through molecular material near the supernova remnant IC 443.  For the 2 sight lines where H$_3^+$ was detected we inferred ionization rates of $\zeta_2=2.6^{+1.3}_{-1.9}\times10^{-15}$~s$^{-1}$ and $\zeta_2=1.6^{+0.8}_{-1.2}\times10^{-15}$~s$^{-1}$---much higher than the mean value reported above---while in the remaining sight lines $3\sigma$ upper limits on $\zeta_2$ were consistent with typical diffuse cloud values.  Not only do these findings support the hypothesis that proximity to a site of particle acceleration controls the cosmic-ray ionization rate, but they also demonstrate variations in $\zeta_2$ on length scales of a few pc.

\section{SUMMARY}

The survey of H$_3^+$ in diffuse molecular clouds now covers 50 sight lines, with detections in 21 of those.  Cosmic-ray ionization rates (and upper limits) are inferred in all of these sight lines.  Where H$_3^+$ is detected, ionization rates are in the range $(1.7\pm1.3)\times10^{-16}$~s$^{-1}<\zeta_2<(10.6\pm8.2)\times10^{-16}$~s$^{-1}$.  Accounting for upper limits, the mean value of the ionization rate is $\zeta_2=(3.5^{+5.3}_{-3.0})\times10^{-16}$~s$^{-1}$.  This is about 1 order of magnitude larger than ionization rates inferred from previous observations of other molecular species (e.g., OH and HD), although recent observations of OH$^+$ and H$_2$O$^+$ with {\it Herschel} are suggestive of high ionization rates as well \citep[$0.6\times10^{-16}$~s$^{-1}<\zeta_{\rm H}< 2.4\times10^{-16}$~s$^{-1}$;][]{neufeld2010}.  The lowest $3\sigma$ upper limits found for sight lines where H$_3^+$ is not detected are about $\zeta_2<0.4\times10^{-16}$~s$^{-1}$.  Together, the wide range of inferred ionization rates and the low upper limits allude to variations in $\zeta_2$ between different diffuse cloud sight lines.

Comparisons of $\zeta_2$ with various line-of-sight properties (including Galactic latitude, Galactic longitude and heliocentric distance) show no clear trends.  This suggests that variations in the cosmic-ray ionization rate between different sight lines are caused not by large scale, but local effects.  A comparison of $\zeta_2$ with the total hydrogen column density ($N_{\rm H}$) shows no strong correlation when only diffuse clouds are considered.  When ionization rates inferred for dense cloud sight lines with much higher $N_{\rm H}$ are included though, $\zeta_2$ is seen to decrease with increased $N_{\rm H}$.  This correlation is expected as low-energy cosmic rays --- those most efficient at ionization --- will lose all of their energy before reaching the interiors of dense clouds.  The lack of a correlation amongst diffuse cloud sight lines suggests that the particles primarily responsible for ionization must be able to completely penetrate diffuse clouds, while the difference in $\zeta_2$ between diffuse and dense clouds indicates that these same particles must be stopped in the outer layers of dense clouds.

Given that total column density --- at least in the case of diffuse clouds --- is unable to explain variations in the ionization rate, the concept of a cosmic-ray spectrum that is uniform throughout the Galaxy must be reconsidered.  Instead, the flux of low-energy (MeV) cosmic rays is most likely controlled by proximity to local sites of particle acceleration.  Our recent observations showing a high ionization rate toward the supernova remnant IC 443 \citep{indriolo2010b} support this conjecture.

\mbox{}

The authors would like to thank Takeshi Oka, Tom Geballe, and Geoff Blake for assistance with observations, Ed Sutton for assistance with treating upper limits in the analysis of the ionization rate distribution, Brian Fields for helpful discussions regarding cosmic rays and particle astrophysics, and the anonymous referee.  N.I. and B.J.M have been supported by NSF grant PHY 08-55633.  Based in part on observations made with ESO Telescopes at the La Silla or Paranal Observatories under programme ID 384.C-0618.  Some of the data presented herein were obtained at the W.M. Keck Observatory, which is operated as a scientific partnership among the California Institute of Technology, the University of California and the National Aeronautics and Space Administration. The Observatory was made possible by the generous financial support of the W.M. Keck Foundation.  The United Kingdom Infrared Telescope is operated by the Joint Astronomy Centre on behalf of the Science and Technology Facilities Council of the U.K.
Based in part on observations obtained at the Gemini Observatory, which is operated by the Association of Universities for Research in Astronomy, Inc., under a cooperative agreement with the NSF on behalf of the Gemini partnership: the National Science Foundation (United States), the Science and Technology Facilities Council (United Kingdom), the National Research Council (Canada), CONICYT (Chile), the Australian Research Council (Australia), Minist\'{e}rio da Ci\^{e}ncia e Tecnologia (Brazil) and Ministerio de Ciencia, Tecnolog\'{i}a e Innovaci\'{o}n Productiva (Argentina).  Gemini/Phoenix spectra were obtained through programs GS-2009B-Q-71 and GS-2010A-Q-60.  This work is also based in part on observations obtained with the Phoenix infrared spectrograph, developed and operated by the National Optical Astronomy Observatory.

%%%%%%%%%%%%%%%%%%%%%%%%%%%%bibliography%%%%%%%%%%%%%%%%%%%%%%%%%%%%%%%%%%%%%%%%%%%%%%%%%
%\bibliographystyle{apj}
%\bibliography{indy_master}

\begin{thebibliography}{}

\bibitem[{{Abdo} {et~al.}(2010){Abdo}, {Ackermann}, {Ajello}, {Baldini},
  {Ballet}, {Barbiellini}, {Bastieri}, {Baughman}, {Bechtol}, {Bellazzini},
  {Berenji}, {Blandford}, {Bloom}, {Bonamente}, {Borgland}, {Bregeon}, {Brez},
  {Brigida}, {Bruel}, {Burnett}, {Buson}, {Caliandro}, {Cameron}, {Caraveo},
  {Casandjian}, {Cecchi}, {{\c C}elik}, {Chekhtman}, {Cheung}, {Chiang},
  {Cillis}, {Ciprini}, {Claus}, {Cohen-Tanugi}, {Cominsky}, {Conrad}, {Cutini},
  {Dermer}, {de Angelis}, {de Palma}, {Silva}, {Drell}, {Drlica-Wagner},
  {Dubois}, {Dumora}, {Farnier}, {Favuzzi}, {Fegan}, {Focke}, {Fortin},
  {Frailis}, {Fukazawa}, {Funk}, {Fusco}, {Gargano}, {Gasparrini}, {Gehrels},
  {Germani}, {Giavitto}, {Giebels}, {Giglietto}, {Giordano}, {Glanzman},
  {Godfrey}, {Grenier}, {Grondin}, {Grove}, {Guillemot}, {Guiriec}, {Hanabata},
  {Harding}, {Hayashida}, {Hughes}, {Jackson}, {J{\'o}hannesson}, {Johnson},
  {Johnson}, {Johnson}, {Kamae}, {Katagiri}, {Kataoka}, {Kawai}, {Kerr},
  {Kn{\"o}dlseder}, {Kocian}, {Kuss}, {Lande}, {Latronico}, {Lee},
  {Lemoine-Goumard}, {Longo}, {Loparco}, {Lott}, {Lovellette}, {Lubrano},
  {Madejski}, {Makeev}, {Mazziotta}, {Meurer}, {Michelson}, {Mitthumsiri},
  {Moiseev}, {Monte}, {Monzani}, {Morselli}, {Moskalenko}, {Murgia},
  {Nakamori}, {Nolan}, {Norris}, {Nuss}, {Ohsugi}, {Orlando}, {Ormes}, {Ozaki},
  {Paneque}, {Panetta}, {Parent}, {Pelassa}, {Pepe}, {Pesce-Rollins}, {Piron},
  {Porter}, {Rain{\`o}}, {Rando}, {Razzano}, {Reimer}, {Reimer}, {Reposeur},
  {Rochester}, {Rodriguez}, {Romani}, {Roth}, {Ryde}, {Sadrozinski}, {Sanchez},
  {Sander}, {Saz Parkinson}, {Scargle}, {Sgr{\`o}}, {Siskind}, {Smith},
  {Smith}, {Spandre}, {Spinelli}, {Strickman}, {Strong}, {Suson}, {Tajima},
  {Takahashi}, {Takahashi}, {Tanaka}, {Thayer}, {Thayer}, {Thompson},
  {Tibaldo}, {Torres}, {Tosti}, {Tramacere}, {Uchiyama}, {Usher}, {Van Etten},
  {Vasileiou}, {Venter}, {Vilchez}, {Vitale}, {Waite}, {Wang}, {Winer}, {Wood},
  {Ylinen}, \& {Ziegler}}]{abdo2010ic443}
{Abdo}, A.~A., {Ackermann}, M., {Ajello}, M., {et~al.} 2010, \apj, 712, 459

\bibitem[{{Acciari} {et~al.}(2009){Acciari}, {Aliu}, {Arlen}, {Aune},
  {Bautista}, {Beilicke}, {Benbow}, {Bradbury}, {Buckley}, {Bugaev}, {Butt},
  {Byrum}, {Cannon}, {Celik}, {Cesarini}, {Chow}, {Ciupik}, {Cogan}, {Colin},
  {Cui}, {Daniel}, {Dickherber}, {Duke}, {Dwarkadas}, {Ergin}, {Fegan},
  {Finley}, {Finnegan}, {Fortin}, {Fortson}, {Furniss}, {Gall}, {Gibbs},
  {Gillanders}, {Godambe}, {Grube}, {Guenette}, {Gyuk}, {Hanna}, {Hays},
  {Holder}, {Horan}, {Hui}, {Humensky}, {Imran}, {Kaaret}, {Karlsson},
  {Kertzman}, {Kieda}, {Kildea}, {Konopelko}, {Krawczynski}, {Krennrich},
  {Lang}, {LeBohec}, {Maier}, {McCann}, {McCutcheon}, {Millis}, {Moriarty},
  {Ong}, {Otte}, {Pandel}, {Perkins}, {Pohl}, {Quinn}, {Ragan}, {Reyes},
  {Reynolds}, {Roache}, {Rose}, {Schroedter}, {Sembroski}, {Smith}, {Steele},
  {Swordy}, {Theiling}, {Toner}, {Valcarcel}, {Varlotta}, {Vassiliev},
  {Vincent}, {Wagner}, {Wakely}, {Ward}, {Weekes}, {Weinstein}, {Weisgarber},
  {Williams}, {Wissel}, {Wood}, \& {Zitzer}}]{acciari2009}
{Acciari}, V.~A., {Aliu}, E., {Arlen}, T., {et~al.} 2009, \apjl, 698, L133

\bibitem[{{Ackermann} {et~al.}(2011)}]{ackermann2011}
{Ackermann}, M., {Ajello}, M., {Allafort}, A., {et~al.} 2011, Science, 334, 1103

\bibitem[{{Albert} {et~al.}(2007){Albert}, {Aliu}, {Anderhub}, {Antoranz},
  {Armada}, {Baixeras}, {Barrio}, {Bartko}, {Bastieri}, {Becker}, {Bednarek},
  {Berger}, {Bigongiari}, {Biland}, {Bock}, {Bordas}, {Bosch-Ramon}, {Bretz},
  {Britvitch}, {Camara}, {Carmona}, {Chilingarian}, {Coarasa}, {Commichau},
  {Contreras}, {Cortina}, {Costado}, {Curtef}, {Danielyan}, {Dazzi}, {De
  Angelis}, {Delgado}, {de los Reyes}, {De Lotto}, {Domingo-Santamar{\'{\i}}a},
  {Dorner}, {Doro}, {Errando}, {Fagiolini}, {Ferenc}, {Fern{\'a}ndez}, {Firpo},
  {Flix}, {Fonseca}, {Font}, {Fuchs}, {Galante}, {Garc{\'{\i}}a-L{\'o}pez},
  {Garczarczyk}, {Gaug}, {Giller}, {Goebel}, {Hakobyan}, {Hayashida},
  {Hengstebeck}, {Herrero}, {H{\"o}hne}, {Hose}, {Hsu}, {Jacon}, {Jogler},
  {Kosyra}, {Kranich}, {Kritzer}, {Laille}, {Lindfors}, {Lombardi}, {Longo},
  {L{\'o}pez}, {L{\'o}pez}, {Lorenz}, {Majumdar}, {Maneva}, {Mannheim},
  {Mansutti}, {Mariotti}, {Mart{\'{\i}}nez}, {Mazin}, {Merck}, {Meucci},
  {Meyer}, {Miranda}, {Mirzoyan}, {Mizobuchi}, {Moralejo}, {Nieto}, {Nilsson},
  {Ninkovic}, {O{\~n}a-Wilhelmi}, {Otte}, {Oya}, {Paneque}, {Panniello},
  {Paoletti}, {Paredes}, {Pasanen}, {Pascoli}, {Pauss}, {Pegna}, {Persic},
  {Peruzzo}, {Piccioli}, {Prandini}, {Puchades}, {Raymers}, {Rhode},
  {Rib{\'o}}, {Rico}, {Rissi}, {Robert}, {R{\"u}gamer}, {Saggion}, {Saito},
  {S{\'a}nchez}, {Sartori}, {Scalzotto}, {Scapin}, {Schmitt}, {Schweizer},
  {Shayduk}, {Shinozaki}, {Shore}, {Sidro}, {Sillanp{\"a}{\"a}}, {Sobczynska},
  {Stamerra}, {Stark}, {Takalo}, {Temnikov}, {Tescaro}, {Teshima}, {Torres},
  {Turini}, {Vankov}, {Vitale}, {Wagner}, {Wibig}, {Wittek}, {Zandanel},
  {Zanin}, \& {Zapatero}}]{albert2007}
{Albert}, J., {Aliu}, E., {Anderhub}, H., {et~al.} 2007, \apjl, 664, L87

\bibitem[{{Black} \& {Dalgarno}(1977)}]{black1977}
{Black}, J.~H., \& {Dalgarno}, A. 1977, \apjs, 34, 405

\bibitem[{{Black} {et~al.}(1978){Black}, {Hartquist}, \&
  {Dalgarno}}]{black1978}
{Black}, J.~H., {Hartquist}, T.~W., \& {Dalgarno}, A. 1978, \apj, 224, 448

\bibitem[{{Bohlin} {et~al.}(1978){Bohlin}, {Savage}, \& {Drake}}]{bohlin1978}
{Bohlin}, R.~C., {Savage}, B.~D., \& {Drake}, J.~F. 1978, \apj, 224, 132

\bibitem[{{Brittain} {et~al.}(2004){Brittain}, {Simon}, {Kulesa}, \&
  {Rettig}}]{brittain2004}
{Brittain}, S.~D., {Simon}, T., {Kulesa}, C., \& {Rettig}, T.~W. 2004, \apj,
  606, 911

\bibitem[{{Cardelli} {et~al.}(1996){Cardelli}, {Meyer}, {Jura}, \&
  {Savage}}]{cardelli1996}
{Cardelli}, J.~A., {Meyer}, D.~M., {Jura}, M., \& {Savage}, B.~D. 1996, \apj,
  467, 334

\bibitem[{{Cartledge} {et~al.}(2004){Cartledge}, {Lauroesch}, {Meyer}, \&
  {Sofia}}]{cartledge2004}
{Cartledge}, S.~I.~B., {Lauroesch}, J.~T., {Meyer}, D.~M., \& {Sofia}, U.~J.
  2004, \apj, 613, 1037

\bibitem[{{Chevalier} \& {Ilovaisky}(1998)}]{chevalier1998}
{Chevalier}, C., \& {Ilovaisky}, S.~A. 1998, \aap, 330, 201

\bibitem[{{Claussen} {et~al.}(1997){Claussen}, {Frail}, {Goss}, \&
  {Gaume}}]{claussen1997}
{Claussen}, M.~J., {Frail}, D.~A., {Goss}, W.~M., \& {Gaume}, R.~A. 1997, \apj,
  489, 143

\bibitem[{{Conti} \& {Vacca}(1990)}]{conti1990}
{Conti}, P.~S., \& {Vacca}, W.~D. 1990, \aj, 100, 431

\bibitem[{{Crabtree} {et~al.}(2011){Crabtree}, {Indriolo}, {Kreckel}, {Tom}, \&
  {McCall}}]{crabtree2011}
{Crabtree}, K.~N., {Indriolo}, N., {Kreckel}, H., {Tom}, B.~A., \& {McCall},
  B.~J. 2011, \apj, 729, 15

\bibitem[{{Cravens} \& {Dalgarno}(1978)}]{cravens1978}
{Cravens}, T.~E., \& {Dalgarno}, A. 1978, \apj, 219, 750

\bibitem[{{Dalgarno}(2006)}]{dalgarno2006}
{Dalgarno}, A. 2006, Proc. Nat. Acad. Sci., 103, 12269

\bibitem[{{de Zeeuw} {et~al.}(1999){de Zeeuw}, {Hoogerwerf}, {de Bruijne},
  {Brown}, \& {Blaauw}}]{dezeeuw1999}
{de Zeeuw}, P.~T., {Hoogerwerf}, R., {de Bruijne}, J.~H.~J., {Brown}, A.~G.~A.,
  \& {Blaauw}, A. 1999, \aj, 117, 354

\bibitem[{{Dickman} {et~al.}(1992){Dickman}, {Snell}, {Ziurys}, \&
  {Huang}}]{dickman1992}
{Dickman}, R.~L., {Snell}, R.~L., {Ziurys}, L.~M., \& {Huang}, Y. 1992, \apj,
  400, 203

\bibitem[{{Federman} {et~al.}(1996){Federman}, {Weber}, \&
  {Lambert}}]{federman1996}
{Federman}, S.~R., {Weber}, J., \& {Lambert}, D.~L. 1996, \apj, 463, 181

\bibitem[{{Fruscione} {et~al.}(1994){Fruscione}, {Hawkins}, {Jelinsky}, \&
  {Wiercigroch}}]{fruscione1994}
{Fruscione}, A., {Hawkins}, I., {Jelinsky}, P., \& {Wiercigroch}, A. 1994,
  \apjs, 94, 127

\bibitem[{{Geballe} {et~al.}(1999){Geballe}, {McCall}, {Hinkle}, \&
  {Oka}}]{geballe1999}
{Geballe}, T.~R., {McCall}, B.~J., {Hinkle}, K.~H., \& {Oka}, T. 1999, \apj,
  510, 251

\bibitem[{{Geballe} \& {Oka}(1996)}]{geballe1996}
{Geballe}, T.~R., \& {Oka}, T. 1996, \nat, 384, 334

\bibitem[{{Geballe} \& {Oka}(2010)}]{geballe2010}
{Geballe}, T.~R., \& {Oka}, T. 2010, \apjl, 709, L70

\bibitem[{{Gerin} {et~al.}(2010){Gerin}, {de Luca}, {Black}, {Goicoechea},
  {Herbst}, {Neufeld}, {Falgarone}, {Godard}, {Pearson}, {Lis}, {Phillips},
  {Bell}, {Sonnentrucker}, {Boulanger}, {Cernicharo}, {Coutens}, {Dartois},
  {Encrenaz}, {Giesen}, {Goldsmith}, {Gupta}, {Gry}, {Hennebelle},
  {Hily-Blant}, {Joblin}, {Kazmierczak}, {Kolos}, {Krelowski},
  {Martin-Pintado}, {Monje}, {Mookerjea}, {Perault}, {Persson}, {Plume},
  {Rimmer}, {Salez}, {Schmidt}, {Stutzki}, {Teyssier}, {Vastel}, {Yu},
  {Contursi}, {Menten}, {Geballe}, {Schlemmer}, {Shipman}, {Tielens},
  {Philipp-May}, {Cros}, {Zmuidzinas}, {Samoska}, {Klein}, \&
  {Lorenzani}}]{gerin2010}
{Gerin}, M., {de Luca}, M., {Black}, J., {et~al.} 2010, \aap, 518, L110

\bibitem[{{Gibb} {et~al.}(2010){Gibb}, {Brittain}, {Rettig}, {Troutman},
  {Simon}, \& {Kulesa}}]{gibb2010}
{Gibb}, E.~L., {Brittain}, S.~D., {Rettig}, T.~W., {et~al.} 2010, \apj, 715, 757

\bibitem[{{Goto} {et~al.}(2002){Goto}, {McCall}, {Geballe}, {Usuda},
  {Kobayashi}, {Terada}, \& {Oka}}]{goto2002}
{Goto}, M., {McCall}, B.~J., {Geballe}, T.~R., {et~al.} 2002, \pasj, 54, 951

\bibitem[{{Goto} {et~al.}(2011){Goto}, {Usuda}, {Geballe}, {Indriolo},
  {McCall}, {Henning}, \& {Oka}}]{goto2011}
{Goto}, M., {Usuda}, T., {Geballe}, T.~R., {et~al.} 2011, PASJ, 63, L13

\bibitem[{{Goto} {et~al.}(2008){Goto}, {Usuda}, {Nagata}, {Geballe}, {McCall},
  {Indriolo}, {Suto}, {Henning}, {Morong}, \& {Oka}}]{goto2008}
{Goto}, M., {Usuda}, T., {Nagata}, T., {et~al.} 2008, \apj, 688, 306

\bibitem[{{Hartquist} {et~al.}(1978{\natexlab{a}}){Hartquist}, {Black}, \&
  {Dalgarno}}]{hartquist1978}
{Hartquist}, T.~W., {Black}, J.~H., \& {Dalgarno}, A. 1978{\natexlab{a}},
  \mnras, 185, 643

\bibitem[{{Hartquist} {et~al.}(1978{\natexlab{b}}){Hartquist}, {Doyle}, \&
  {Dalgarno}}]{hartquist1978a}
{Hartquist}, T.~W., {Doyle}, H.~T., \& {Dalgarno}, A. 1978{\natexlab{b}}, \aap,
  68, 65

\bibitem[{{Hayakawa} {et~al.}(1961){Hayakawa}, {Nishimura}, \&
  {Takayanagi}}]{hayakawa1961}
{Hayakawa}, S., {Nishimura}, S., \& {Takayanagi}, T. 1961, \pasj, 13, 184

\bibitem[{{Herbst} \& {Klemperer}(1973)}]{herbst1973}
{Herbst}, E., \& {Klemperer}, W. 1973, \apj, 185, 505

\bibitem[{{Hewitt} {et~al.}(2006){Hewitt}, {Yusef-Zadeh}, {Wardle}, {Roberts},
  \& {Kassim}}]{hewitt2006}
{Hewitt}, J.~W., {Yusef-Zadeh}, F., {Wardle}, M., {Roberts}, D.~A., \&
  {Kassim}, N.~E. 2006, \apj, 652, 1288

\bibitem[{{Hezareh} {et~al.}(2008){Hezareh}, {Houde}, {McCoey}, {Vastel}, \&
  {Peng}}]{hezareh2008}
{Hezareh}, T., {Houde}, M., {McCoey}, C., {Vastel}, C., \& {Peng}, R. 2008,
  \apj, 684, 1221

\bibitem[{{Hinkle} {et~al.}(2003){Hinkle}, {Blum}, {Joyce}, {Sharp}, {Ridgway},
  {Bouchet}, {van der Bliek}, {Najita}, \& {Winge}}]{hinkle2003}
{Hinkle}, K.~H., {Blum}, R.~D., {Joyce}, R.~R., {et~al.} 2003, \procspie, 4834, 353

\bibitem[{{Huang} {et~al.}(1986){Huang}, {Dickman}, \& {Snell}}]{huang1986}
{Huang}, Y., {Dickman}, R.~L., \& {Snell}, R.~L. 1986, \apjl, 302, L63

\bibitem[{{Indriolo}(2011)}]{indriolo2011}
{Indriolo}, N. 2011, PhD thesis, University of Illinois

\bibitem[{{Indriolo} {et~al.}(2010){Indriolo}, {Blake}, {Goto}, {Usuda}, {Oka},
  {Geballe}, {Fields}, \& {McCall}}]{indriolo2010b}
{Indriolo}, N., {Blake}, G.~A., {Goto}, M., {et~al.} 2010, \apj, 724, 1357

\bibitem[{{Indriolo} {et~al.}(2007){Indriolo}, {Geballe}, {Oka}, \&
  {McCall}}]{indriolo2007}
{Indriolo}, N., {Geballe}, T.~R., {Oka}, T., \& {McCall}, B.~J. 2007, \apj,
  671, 1736

\bibitem[{{Jenkins} {et~al.}(1983){Jenkins}, {Jura}, \&
  {Loewenstein}}]{jenkins1983}
{Jenkins}, E.~B., {Jura}, M., \& {Loewenstein}, M. 1983, \apj, 270, 88

\bibitem[{{Jensen} {et~al.}(2005){Jensen}, {Rachford}, \& {Snow}}]{jensen2005}
{Jensen}, A.~G., {Rachford}, B.~L., \& {Snow}, T.~P. 2005, \apj, 619, 891

\bibitem[{{Jura}(1975)}]{jura1975}
{Jura}, M. 1975, \apj, 197, 581

\bibitem[{{Karpas} {et~al.}(1979){Karpas}, {Anicich}, \&
  {Huntress}}]{karpas1979}
{Karpas}, Z., {Anicich}, V., \& {Huntress}, W.~T. 1979, J. Chem. Phys., 70,
  2877

\bibitem[{{K\"{a}ufl} {et~al.}(2004){K\"{a}ufl}, {Ballester}, {Biereichel},
  {Delabre}, {Donaldson}, {Dorn}, {Fedrigo}, {Finger}, {Fischer}, {Franza},
  {Gojak}, {Huster}, {Jung}, {Lizon}, {Mehrgan}, {Meyer}, {Moorwood}, {Pirard},
  {Paufique}, {Pozna}, {Siebenmorgen}, {Silber}, {Stegmeier}, \&
  {Wegerer}}]{kaufl2004}
{K\"{a}ufl}, H., {Ballester}, P., {Biereichel}, P., {et~al.} 2004, \procspie, 5492, 1218

\bibitem[{Klippenstein {et~al.}(2010)Klippenstein, Georgievskii, \&
  McCall}]{klippenstein2010}
Klippenstein, S.~J., Georgievskii, Y., \& McCall, B.~J. 2010, J. Phys. Chem. A,
  114, 278

\bibitem[{{Kreckel} {et~al.}(2005){Kreckel}, {Motsch}, {Mikosch},
  {Glos{\'{\i}}k}, {Pla{\v s}il}, {Altevogt}, {Andrianarijaona}, {Buhr},
  {Hoffmann}, {Lammich}, {Lestinsky}, {Nevo}, {Novotny}, {Orlov}, {Pedersen},
  {Sprenger}, {Terekhov}, {Toker}, {Wester}, {Gerlich}, {Schwalm}, {Wolf}, \&
  {Zajfman}}]{kreckel2005}
{Kreckel}, H., {Motsch}, M., {Mikosch}, J., {et~al.} 2005, Physical Review Letters, 95, 263201

\bibitem[{{Kreckel} {et~al.}(2010){Kreckel}, {Novotn{\'y}}, {Crabtree}, {Buhr},
  {Petrignani}, {Tom}, {Thomas}, {Berg}, {Bing}, {Grieser}, {Krantz},
  {Lestinsky}, {Mendes}, {Nordhorn}, {Repnow}, {St{\"u}tzel}, {Wolf}, \&
  {McCall}}]{kreckel2010}
{Kreckel}, H., {Novotn{\'y}}, O., {Crabtree}, K.~N., {et~al.} 2010, \pra, 82, 042715

\bibitem[{{Kulesa}(2002)}]{kulesa2002}
{Kulesa}, C.~A. 2002, PhD thesis, The University of Arizona

\bibitem[{{Lodders}(2003)}]{lodders2003}
{Lodders}, K. 2003, \apj, 591, 1220

\bibitem[{{Ma{\'{\i}}z-Apell{\'a}niz}
  {et~al.}(2004){Ma{\'{\i}}z-Apell{\'a}niz}, {Walborn}, {Galu{\'e}}, \&
  {Wei}}]{maiz2004}
{Ma{\'{\i}}z-Apell{\'a}niz}, J., {Walborn}, N.~R., {Galu{\'e}}, H.~{\'A}., \&
  {Wei}, L.~H. 2004, \apjs, 151, 103

\bibitem[{{McCall}(2001)}]{mccall2001}
{McCall}, B.~J. 2001, PhD thesis, The University of Chicago

\bibitem[{{McCall} {et~al.}(2010){McCall}, {Drosback}, {Thorburn}, {York},
  {Friedman}, {Hobbs}, {Rachford}, {Snow}, {Sonnentrucker}, \&
  {Welty}}]{mccall2010}
{McCall}, B.~J., {Drosback}, M.~M., {Thorburn}, J.~A., {et~al.} 2010, \apj, 708, 1628

\bibitem[{{McCall} {et~al.}(1998){McCall}, {Geballe}, {Hinkle}, \&
  {Oka}}]{mccall1998}
{McCall}, B.~J., {Geballe}, T.~R., {Hinkle}, K.~H., \& {Oka}, T. 1998, Science,
  279, 1910

\bibitem[{{McCall} {et~al.}(1999){McCall}, {Geballe}, {Hinkle}, \&
  {Oka}}]{mccall1999}
{McCall}, B.~J., {Geballe}, T.~R., {Hinkle}, K.~H., \& {Oka}, T. 1999, \apj, 522, 338

\bibitem[{{McCall} {et~al.}(2002){McCall}, {Hinkle}, {Geballe},
  {Moriarty-Schieven}, {Evans}, {Kawaguchi}, {Takano}, {Smith}, \&
  {Oka}}]{mccall2002}
{McCall}, B.~J., {Hinkle}, K.~H., {Geballe}, T.~R., {et~al.} 2002, \apj, 567, 391

\bibitem[{{McCall} {et~al.}(2003){McCall}, {Huneycutt}, {Saykally}, {Geballe},
  {Djuric}, {Dunn}, {Semaniak}, {Novotny}, {Al-Khalili}, {Ehlerding},
  {Hellberg}, {Kalhori}, {Neau}, {Thomas}, {{\"O}sterdahl}, \&
  {Larsson}}]{mccall2003}
{McCall}, B.~J., {Huneycutt}, A.~J., {Saykally}, R.~J., {et~al.} 2003, \nat, 422, 500

\bibitem[{{McCall} {et~al.}(2004){McCall}, {Huneycutt}, {Saykally}, {Djuric},
  {Dunn}, {Semaniak}, {Novotny}, {Al-Khalili}, {Ehlerding}, {Hellberg},
  {Kalhori}, {Neau}, {Thomas}, {Paal}, {{\"O}sterdahl}, \&
  {Larsson}}]{mccall2004}
{McCall}, B.~J., {Huneycutt}, A.~J., {Saykally}, R.~J., {et~al.} 2004, Phys. Rev. A, 70, 052716

\bibitem[{{McLean} {et~al.}(1998){McLean}, {Becklin}, {Bendiksen}, {Brims},
  {Canfield}, {Figer}, {Graham}, {Hare}, {Lacayanga}, {Larkin}, {Larson},
  {Levenson}, {Magnone}, {Teplitz}, \& {Wong}}]{mclean1998}
{McLean}, I.~S., {Becklin}, E.~E., {Bendiksen}, O., {et~al.} 1998, \procspie, 3354, 566

\bibitem[{{Mitchell}(1990)}]{mitchell1990}
{Mitchell}, J.~B.~A. 1990, \physrep, 186, 215

\bibitem[{{Mountain} {et~al.}(1990){Mountain}, {Robertson}, {Lee}, \&
  {Wade}}]{mountain1990}
{Mountain}, C.~M., {Robertson}, D.~J., {Lee}, T.~J., \& {Wade}, R. 1990,
  \procspie, 1235, 25

\bibitem[{{Neufeld} {et~al.}(2010){Neufeld}, {Goicoechea}, {Sonnentrucker},
  {Black}, {Pearson}, {Yu}, {Phillips}, {Lis}, {de Luca}, {Herbst}, {Rimmer},
  {Gerin}, {Bell}, {Boulanger}, {Cernicharo}, {Coutens}, {Dartois},
  {Kazmierczak}, {Encrenaz}, {Falgarone}, {Geballe}, {Giesen}, {Godard},
  {Goldsmith}, {Gry}, {Gupta}, {Hennebelle}, {Hily-Blant}, {Joblin},
  {Ko{\l}os}, {Kre{\l}owski}, {Mart{\'{\i}}n-Pintado}, {Menten}, {Monje},
  {Mookerjea}, {Perault}, {Persson}, {Plume}, {Salez}, {Schlemmer}, {Schmidt},
  {Stutzki}, {Teyssier}, {Vastel}, {Cros}, {Klein}, {Lorenzani}, {Philipp},
  {Samoska}, {Shipman}, {Tielens}, {Szczerba}, \& {Zmuidzinas}}]{neufeld2010}
{Neufeld}, D.~A., {Goicoechea}, J.~R., {Sonnentrucker}, P., {et~al.} 2010, \aap, 521, L10

\bibitem[{{Nichols} \& {Slavin}(2004)}]{nichols2004}
{Nichols}, J.~S., \& {Slavin}, J.~D. 2004, \apj, 610, 285

\bibitem[{{Nota} {et~al.}(1996){Nota}, {Pasquali}, {Clampin}, {Pollacco},
  {Scuderi}, \& {Livio}}]{nota1996}
{Nota}, A., {Pasquali}, A., {Clampin}, M., {et~al.} 1996, \apj, 473, 946

\bibitem[{{O'Donnell} \& {Watson}(1974)}]{odonnell1974}
{O'Donnell}, E.~J., \& {Watson}, W.~D. 1974, \apj, 191, 89

\bibitem[{{Oka} {et~al.}(2005){Oka}, {Geballe}, {Goto}, {Usuda}, \&
  {McCall}}]{oka2005}
{Oka}, T., {Geballe}, T.~R., {Goto}, M., {Usuda}, T., \& {McCall}, B.~J. 2005,
  \apj, 632, 882

\bibitem[{{Padovani} {et~al.}(2009){Padovani}, {Galli}, \&
  {Glassgold}}]{padovani2009}
{Padovani}, M., {Galli}, D., \& {Glassgold}, A.~E. 2009, \aap, 501, 619

\bibitem[{{Pasquali} {et~al.}(2002){Pasquali}, {Nota}, {Smith}, {Akiyama},
  {Messineo}, \& {Clampin}}]{pasquali2002}
{Pasquali}, A., {Nota}, A., {Smith}, L.~J., {et~al.} 2002, \aj, 124, 1625

\bibitem[{{Rachford} {et~al.}(2009){Rachford}, {Snow}, {Destree}, {Ross},
  {Ferlet}, {Friedman}, {Gry}, {Jenkins}, {Morton}, {Savage}, {Shull},
  {Sonnentrucker}, {Tumlinson}, {Vidal-Madjar}, {Welty}, \&
  {York}}]{rachford2009}
{Rachford}, B.~L., {Snow}, T.~P. and {Destree}, J.~D., {et~al.} 2009, \apjs, 180, 125

\bibitem[{{Rachford} {et~al.}(2002){Rachford}, {Snow}, {Tumlinson}, {Shull},
  {Blair}, {Ferlet}, {Friedman}, {Gry}, {Jenkins}, {Morton}, {Savage},
  {Sonnentrucker}, {Vidal-Madjar}, {Welty}, \& {York}}]{rachford2002}
{Rachford}, B.~L., {Snow}, T.~P., {Tumlinson}, J., {et~al.} 2002, \apj, 577, 221

\bibitem[{Rakshit(1982)}]{rakshit1982}
Rakshit, A.~B. 1982, International Journal or Mass Spectrometry and Ion
  Physics, 41, 185

\bibitem[{{Ritchey} {et~al.}(2011){Ritchey}, {Federman}, \&
  {Lambert}}]{ritchey2011}
{Ritchey}, A.~M., {Federman}, S.~R., \& {Lambert}, D.~L. 2011, \apj, 728, 36

\bibitem[{{Roberts} {et~al.}(2010){Roberts}, {Gies}, {Parks}, {Grundstrom},
  {McSwain}, {Berger}, {Mason}, {ten Brummelaar}, \& {Turner}}]{roberts2010}
{Roberts}, Jr., L.~C., {Gies}, D.~R., {Parks}, J.~R., {et~al.} 2010, \aj, 140, 744

\bibitem[{{Savage} {et~al.}(1977){Savage}, {Bohlin}, {Drake}, \&
  {Budich}}]{savage1977}
{Savage}, B.~D., {Bohlin}, R.~C., {Drake}, J.~F., \& {Budich}, W. 1977, \apj,
  216, 291

\bibitem[{{Sheffer} {et~al.}(2008){Sheffer}, {Rogers}, {Federman}, {Abel},
  {Gredel}, {Lambert}, \& {Shaw}}]{sheffer2008}
{Sheffer}, Y., {Rogers}, M., {Federman}, S.~R., {et~al.} 2008, \apj, 687, 1075

\bibitem[{{Shuping} {et~al.}(1999){Shuping}, {Snow}, {Crutcher}, \&
  {Lutz}}]{shuping1999}
{Shuping}, R.~Y., {Snow}, T.~P., {Crutcher}, R., \& {Lutz}, B.~L. 1999, \apj,
  520, 149

\bibitem[{{Snow} {et~al.}(2010){Snow}, {Destree}, {Burgh}, {Ferguson},
  {Danforth}, \& {Cordiner}}]{snow2010}
{Snow}, T.~P., {Destree}, J.~D., {Burgh}, E.~B., {et~al.} 2010, \apjl, 720, L190

\bibitem[{{Snow} \& {McCall}(2006)}]{snow2006}
{Snow}, T.~P., \& {McCall}, B.~J. 2006, ARA\&A, 44, 367

\bibitem[{{Sofia} {et~al.}(2004){Sofia}, {Lauroesch}, {Meyer}, \&
  {Cartledge}}]{sofia2004}
{Sofia}, U.~J., {Lauroesch}, J.~T., {Meyer}, D.~M., \& {Cartledge}, S.~I.~B.
  2004, \apj, 605, 272

\bibitem[{{Sonnentrucker} {et~al.}(2007){Sonnentrucker}, {Welty}, {Thorburn},
  \& {York}}]{sonnentrucker2007}
{Sonnentrucker}, P., {Welty}, D.~E., {Thorburn}, J.~A., \& {York}, D.~G. 2007,
  \apjs, 168, 58

\bibitem[{{Spitzer} \& {Tomasko}(1968)}]{spitzer1968}
{Spitzer}, Jr., L., \& {Tomasko}, M.~G. 1968, \apj, 152, 971

\bibitem[{{Tavani} {et~al.}(2010){Tavani}, {Giuliani}, {Chen}, {Argan},
  {Barbiellini}, {Bulgarelli}, {Caraveo}, {Cattaneo}, {Cocco}, {Contessi},
  {D'Ammando}, {Costa}, {De Paris}, {Del Monte}, {Di Cocco}, {Donnarumma},
  {Evangelista}, {Ferrari}, {Feroci}, {Fuschino}, {Galli}, {Gianotti},
  {Labanti}, {Lapshov}, {Lazzarotto}, {Lipari}, {Longo}, {Marisaldi},
  {Mastropietro}, {Mereghetti}, {Morelli}, {Moretti}, {Morselli}, {Pacciani},
  {Pellizzoni}, {Perotti}, {Piano}, {Picozza}, {Pilia}, {Pucella}, {Prest},
  {Rapisarda}, {Rappoldi}, {Scalise}, {Rubini}, {Sabatini}, {Striani},
  {Soffitta}, {Trifoglio}, {Trois}, {Vallazza}, {Vercellone}, {Vittorini},
  {Zambra}, {Zanello}, {Pittori}, {Verrecchia}, {Santolamazza}, {Giommi},
  {Colafrancesco}, {Antonelli}, \& {Salotti}}]{tavani2010}
{Tavani}, M., {Giuliani}, A., {Chen}, A.~W., {et~al.} 2010, \apjl, 710, L151

\bibitem[{{Theard} \& {Huntress}(1974)}]{theard1974}
{Theard}, L.~P., \& {Huntress}, W.~T. 1974, J. Chem. Phys., 60, 2840

\bibitem[{{Thorburn} {et~al.}(2003){Thorburn}, {Hobbs}, {McCall}, {Oka},
  {Welty}, {Friedman}, {Snow}, {Sonnentrucker}, \& {York}}]{thorburn2003}
{Thorburn}, J.~A., {Hobbs}, L.~M., {McCall}, B.~J., {et~al.} 2003, \apj, 584, 339

\bibitem[{{Torres-Dodgen} {et~al.}(1991){Torres-Dodgen}, {Carroll}, \&
  {Tapia}}]{torresdodgen1991}
{Torres-Dodgen}, A.~V., {Carroll}, M., \& {Tapia}, M. 1991, \mnras, 249, 1

\bibitem[{{Tuthill} {et~al.}(2008){Tuthill}, {Monnier}, {Lawrance}, {Danchi},
  {Owocki}, \& {Gayley}}]{tuthill2008}
{Tuthill}, P.~G., {Monnier}, J.~D., {Lawrance}, N., {et~al.} 2008, \apj, 675, 698

\bibitem[{{van der Tak} \& {van Dishoeck}(2000)}]{vandertak2000}
{van der Tak}, F.~F.~S., \& {van Dishoeck}, E.~F. 2000, \aap, 358, L79

\bibitem[{{van Dishoeck} \& {Black}(1986)}]{vandishoeck1986}
{van Dishoeck}, E.~F., \& {Black}, J.~H. 1986, \apjs, 62, 109

\bibitem[{{van Dishoeck} \& {Black}(1989)}]{vandishoeck1989}
{van Dishoeck}, E.~F., \& {Black}, J.~H. 1989, \apj, 340, 273

\bibitem[{{van Leeuwen}(2007)}]{vanleeuwen2007}
{van Leeuwen}, F. 2007, \aap, 474, 653

\bibitem[{Watson(1973)}]{watson1973}
Watson, W.~D. 1973, \apj, 183, L17

\bibitem[{{Webber}(1998)}]{webber1998}
{Webber}, W.~R. 1998, \apj, 506, 329

\end{thebibliography}

%%%%%%%%%%%%%%%%%%%%%%%%%%%%%%tables%%%%%%%%%%%%%%%%%%%%%%%%%%%%%%%%%%%%%%%%%%%%%%%%%%%%

%%%%%%%%%%%%%%%%%%%%%%%%%target table%%%%%%%%%%%%%%%%%%%%%%%%%%

\clearpage
\footnotesize
%\begin{landscape}
\begin{longtable}{lccccc}
%header to appear on first page of table
\caption{Science Targets}\\
           &                  & $E(B-V)$ &       & Distance &      \\
Object & HD number & (mag.)   & Ref. & (pc)       & Ref. \\
\hline
\endfirsthead
%header to appear on subsequent pages of table
\caption[]{(continued)}\\
           &                  & $E(B-V)$ &       & Distance &      \\
Object & HD number & (mag.)   & Ref. & (pc)       & Ref. \\
\hline
\endhead
\hline
\multicolumn{6}{p{4in}}{
{\bf References:} 
%(1) \citet{hirschauer2009}; 
 (1) \citet{mccall2010}; (2) \citet{thorburn2003}; (3) \citet{savage1977}; (4) \citet{vandishoeck1989}; (5) \citet{rachford2009};
(6) \citet{conti1990}; (7) \citet[][and references therein]{mccall2002};
(8) \citet{nota1996}; (9) \citet{shuping1999}; (10) \citet{snow2010}; (11) \citet{vanleeuwen2007};
(12) \citet{fruscione1994}; (13) \citet{nichols2004}; (14) \citet{chevalier1998}; (15) \citet{ritchey2011}; (16) \citet{maiz2004}; (17) \citet{tuthill2008}; (18) \citet{pasquali2002}; (19) \citet{roberts2010}; (20) \citet{sheffer2008}}
\endlastfoot
$\kappa$ Cas &  2905	      & 0.33 &  1 & 1370 & 11 \\
HD 20041       &  20041    & 0.72 &  2 & 1400 &  2 \\
HD 21483       &  21483    & 0.56 &  2 & 440  &  2 \\
HD 21389       &  21389    & 0.57 &  2 & 940  &  2 \\
HD 21856       &  21856    & 0.19 &  3 & 502  & 11 \\
40 Per	           &  22951    & 0.27 &  1 & 283  & 11 \\
$o$ Per     	   &  23180    & 0.31 &  2 & 280  &  2 \\
BD +31 643    	&  281159 & 0.85 &  2 & 240  &  2 \\
$\zeta$ Per  	&  24398   & 0.31 &  2 & 301  &  2 \\
X Per               &  24534    & 0.59 &  2 & 590  &  2 \\
$\epsilon$ Per	&  24760   & 0.09 &  3 & 196  & 11 \\
$\xi$ Per   	    &  24912   & 0.33 &  2 & 470  &  2 \\
62 Tau	    	    &  27778   & 0.37 &  2 & 223  &  2 \\
HD 29647        &  29647   & 1.00 &  2 & 177  &  2 \\
$\alpha$ Cam	&  30614   & 0.30 &  2 & 820  &  2 \\
NGC 2024 IRS 1&  ...         & 1.69 &  4 & 450  &  4 \\
$\chi^2$ Ori	&  41117   & 0.45 &  2 & 1000 &  2 \\
HD 47129    	&  47129   & 0.36 &  1 & 1514 & 12 \\
HD 53367    	&  53367    & 0.74 &  2 & 780  &  2 \\
HD 73882    	&  73882    & 0.70 &  5 & 1000 & 13 \\
HD 110432   	&  110432  & 0.51 &  5 & 300  & 14 \\
$o$ Sco	        &  147084  & 0.74 &  4 & 270  & 11 \\
$\rho$ Oph D 	&  147888  & 0.47 &  2 & 136  &  2 \\
HD 147889   	&  147889   & 1.07 &  2 & 136  &  2 \\
$\chi$ Oph  	&  148184   & 0.52 &  1 & 161  &  5 \\
$\mu$ Nor   	&  149038   & 0.38 &  3 & 1122 & 12 \\
HD 149404   	&  149404   & 0.68 &  1 & 417  & 11 \\
$\zeta$ Oph 	&  149757   & 0.32 &  2 & 140  &  2 \\
HD 152236   	&  152236   & 0.68 &  5 & 600  & 15 \\
HD 154368   	&  154368   & 0.78 &  5 & 909  & 16 \\
WR 104            &  ...	           & 2.30 &  6 & 2600 & 17 \\
HD 168607   	&  168607   & 1.61 &  7 & 1100 &  7 \\
HD 168625   	&  168625  & 1.86 &   8 & 2800 & 18 \\
BD -14 5037 	&  ...	           & 1.57 &  4 & 1700 &  4 \\
HD 169454   	&  169454   & 1.12 &  2 & 930  &  2 \\
W 40 IRS 1A  	&  ...            & 2.90 &  9 & 400  & 9 \\
WR 118            &  ...            & 4.57 &  6 & 3900 &  6 \\
WR 121            &  ...            & 1.75 &  6 & 2600 &  6 \\
HD 183143     	 &  183143  & 1.28 &  7 & 1000 &  7 \\
P Cygni	    	     &  193237  & 0.63 &  7 & 1800 &  7 \\
HD 193322A  	 &  193322A& 0.40 &  3 & 830  & 19 \\
HD 229059   	 &  229059  & 1.71 &  2 & 1000 &  2 \\
HD 194279   	 &  194279  & 1.22 &  7 & 1100 &  7 \\
Cyg OB2 5   	     &  ...	       & 1.99 &  7  & 1700 &  7 \\
Cyg OB2 12  	 &  ...           & 3.35 &  7 & 1700 &  7 \\
Cyg OB2 8A  	 &  ...           & 1.60 & 10 & 1700 &  7 \\
HD 204827       &  204827  & 1.11 &  2 & 600  &  2 \\
HD 206267       &  206267  & 0.53 &  2 & 1000 &  2 \\
19 Cep              &  209975  & 0.36 &  1 & 1300 & 20 \\
$\lambda$ Cep &  210839  & 0.57 &  2 & 505  &  2 \\
1 Cas                &  218376  & 0.25 &  2 & 339  & ~2
\label{table_targets}
\end{longtable}
%\end{landscape}
\normalsize

%%%%%%%%%%%%%%%%%%%%%%%%%%absorption line parameters%%%%%%%%%%%%%%
\clearpage
\scriptsize
%\begin{landscape}
\begin{longtable}{cccccccc}
\caption{Absorption Line Parameters}\\
       &            & $v_{\rm LSR}$ & FWHM          & $W_\lambda$        & $\sigma(W_\lambda)$ & $N(J,K)$              &
$\sigma(N)$ \\
Object & Transition & (km~s$^{-1}$) & (km~s$^{-1}$) & ($10^{-6}$ $\mu$m) & ($10^{-6}$ $\mu$m)  & ($10^{13}$ cm$^{-2}$) &
($10^{13}$ cm$^{-2}$) \\
\hline
\endfirsthead
\caption[]{(continued)}\\
       &            & $v_{\rm LSR}$ & FWHM          & $W_\lambda$        & $\sigma(W_\lambda)$ & $N(J,K)$              &
$\sigma(N)$ \\
Object & Transition & (km~s$^{-1}$) & (km~s$^{-1}$) & ($10^{-6}$ $\mu$m) & ($10^{-6}$ $\mu$m)  & ($10^{13}$ cm$^{-2}$) &
($10^{13}$ cm$^{-2}$) \\
\hline
\endhead
\hline
\multicolumn{8}{p{6in}}{{\bf Notes:} Columns 6 ($\sigma(W_\lambda)$) and 8 ($\sigma(N)$) are $1\sigma$ uncertainties on the equivalent width and column density, respectively.  The FWHM values in the case of non-detections are appropriate for the resolution of the instrument used in each observation, and are used in computing the value of ${\cal N}_{\rm pix}$ required in equation (\ref{eq_eqwid_ul}).}
\endlastfoot
HD 20041      & $R(1,1)^u$ &  2.5 & 11.0 & 2.49 & 0.45 & 10.3 & 1.87 \\
              & $R(1,0)$    & -0.9 & 10.1 & 3.03 & 0.41 & 7.67 & 1.04 \\
HD 21389      & $R(1,1)^u$ &  0.2 & 13.3 & 1.20 & 0.19 & 4.98 & 0.79 \\
              & $R(1,0)$   &  0.8 & 12.9 & 1.46 & 0.19 & 3.69 & 0.47 \\
$\zeta$ Per   & $R(1,1)^u$ &  8.0 & 10.3 & 0.99 & 0.11 & 4.09 & 0.46 \\
              & $R(1,0)$   &  8.4 &  8.7 & 0.86 & 0.10 & 2.17 & 0.25 \\
              & $R(3,3)^l$ &  ... &   10 & ...  & 0.10 & ...  & 0.37 \\
X Per         & $R(1,1)^u$ &  7.0 & 11.7 & 1.16 & 0.20 & 4.81 & 0.81 \\
              & $R(1,0)$   &  6.3 &  9.4 & 1.00 & 0.17 & 2.53 & 0.43 \\
HD 29647      & $R(1,1)^u$ &  8.0 &  8.6 & 1.62 & 0.21 & 6.72 & 0.88 \\
              & $R(1,0)$   &  8.2 & 11.0 & 2.06 & 0.26 & 5.21 & 0.67 \\
HD 73882      & $R(1,1)^u$ &  5.9 &  3.9 & 1.44 & 0.21 & 5.97 & 0.86 \\
              & $R(1,0)$   &  5.7 &  3.2 & 1.16 & 0.19 & 2.94 & 0.48 \\
              & $R(1,1)^l$ &  5.4 &  3.5 & 1.34 & 0.15 & 6.15 & 0.69 \\
HD 110432     & $R(1,1)^u$ & -3.8 &  6.9 & 0.74 & 0.06 & 3.08 & 0.24 \\
              & $R(1,0)$   & -3.3 &  7.5 & 0.83 & 0.07 & 2.11 & 0.17 \\
              & $R(1,1)^l$ & -3.1 &  8.1 & 0.69 & 0.06 & 3.15 & 0.28 \\
HD 154368     & $R(1,1)^u$ &  5.4 &  5.8 & 1.57 & 0.28 & 6.51 & 1.16 \\
              & $R(1,0)$   &  4.6 &  5.1 & 1.13 & 0.25 & 2.86 & 0.63 \\
HD 168607     & $R(1,1)^u$ & 20.9 &  7.9 & 1.07 & 0.13 & 4.44 & 0.53 \\
              & $R(1,0)$   & 22.4 &  8.9 & 0.85 & 0.14 & 2.16 & 0.34 \\
HD 168625     & $R(1,1)^u$ & 19.5 &  9.5 & 1.58 & 0.19 & 6.55 & 0.78 \\
              & $R(1,0)$   & 22.0 &  9.1 & 1.59 & 0.19 & 4.02 & 0.48 \\
HD 169454     & $R(1,1)^u$ &  4.5 & 10.6 & 0.80 & 0.12 & 3.32 & 0.51 \\
              & $R(1,0)$   &  4.0 & 11.7 & 1.01 & 0.12 & 2.56 & 0.30 \\
              & $R(1,1)^l$ &  3.8 &  8.7 & 0.77 & 0.18 & 3.53 & 0.84 \\
W40 IRS 1A    & $R(1,1)^u$ &  5.9 &  8.9 & 4.19 & 0.64 & 17.4 & 2.66 \\
              & $R(1,0)$   &  7.4 &  7.8 & 4.48 & 0.60 & 11.3 & 1.51 \\
              & $R(1,1)^l$ &  6.1 &  4.5 & 2.80 & 0.71 & 12.8 & 3.26 \\
HD 229059     & $R(1,1)^u$ &  3.8 &  9.3 & 5.10 & 0.25 & 21.2 & 1.02 \\
              & $R(1,0)$   &  3.5 &  9.7 & 6.42 & 0.26 & 16.2 & 0.65 \\
HD 204827     & $R(1,1)^u$ & -0.8 & 11.0 & 3.11 & 0.53 & 12.9 & 2.18 \\
              & $R(1,0)$   & -0.9 &  8.7 & 2.41 & 0.51 & 6.10 & 1.29 \\
$\lambda$ Cep & $R(1,1)^u$ & -2.9 & 10.7 & 1.04 & 0.23 & 4.31 & 0.95 \\
              & $R(1,0)$   & -1.4 & 13.5 & 1.29 & 0.27 & 3.26 & 0.67 \\
$\kappa$ Cas  & $R(1,1)^u$ &  ... &   15 & ...  & 0.19 &  ... & 0.77 \\
              & $R(1,0)$   &  ... &   15 & ...  & 0.19 &  ... & 0.47 \\
HD 21483      & $R(1,1)^u$ &  ... &   15 & ...  & 0.47 &  ... & 1.97 \\
              & $R(1,0)$   &  ... &   15 & ...  & 0.47 &  ... & 1.20 \\
HD 21856      & $R(1,1)^u$ &  ... &   15 & ...  & 0.46 &  ... & 1.92 \\
              & $R(1,0)$   &  ... &   15 & ...  & 0.46 &  ... & 1.17 \\
40 Per        & $R(1,1)^u$ &  ... &   15 & ...  & 0.18 &  ... & 0.74 \\
              & $R(1,0)$   &  ... &   15 & ...  & 0.18 &  ... & 0.45 \\
$o$ Per       & $R(1,1)^u$ &  ... &   10 & ...  & 0.10 &  ... & 0.41 \\
              & $R(1,0)$   &  ... &   10 & ...  & 0.10 &  ... & 0.25 \\
BD +31 643    & $R(1,1)^u$ &  ... &   15 & ...  & 0.95 &  ... & 3.95 \\
              & $R(1,0)$   &  ... &   15 & ...  & 0.95 &  ... & 2.41 \\
$\epsilon$ Per& $R(1,1)^u$ &  ... &   10 & ...  & 0.08 &  ... & 0.33 \\
              & $R(1,0)$   &  ... &   10 & ...  & 0.08 &  ... & 0.20 \\
$\xi$ Per     & $R(1,1)^u$ &  ... &   10 & ...  & 0.11 &  ... & 0.44 \\
              & $R(1,0)$   &  ... &   10 & ...  & 0.11 &  ... & 0.27 \\
62 Tau        & $R(1,1)^u$ &  ... &   15 & ...  & 0.37 &  ... & 1.53 \\
              & $R(1,0)$   &  ... &   15 & ...  & 0.37 &  ... & 0.93 \\
$\alpha$ Cam  & $R(1,1)^u$ &  ... &   15 & ...  & 0.30 &  ... & 1.26 \\
              & $R(1,0)$   &  ... &   15 & ...  & 0.30 &  ... & 0.77 \\
NGC 2024 IRS1 & $R(1,1)^u$ &  ... &   10 & ...  & 0.26 &  ... & 1.09 \\
              & $R(1,0)$   &  ... &   10 & ...  & 0.26 &  ... & 0.66 \\
HD 47129      & $R(1,1)^u$ &  ... &   15 & ...  & 0.60 &  ... & 2.48 \\
              & $R(1,0)$   &  ... &   15 & ...  & 0.60 &  ... & 1.51 \\
HD 53367      & $R(1,1)^u$ &  ... &    5 & ...  & 0.12 &  ... & 0.50 \\
              & $R(1,0)$   &  ... &    5 & ...  & 0.12 &  ... & 0.30 \\
              & $R(1,1)^l$ &  ... &    5 & ...  & 0.07 &  ... & 0.32 \\
$o$ Sco       & $R(1,1)^u$ &  ... &   10 & ...  & 0.08 &  ... & 0.35 \\
              & $R(1,0)$   &  ... &   10 & ...  & 0.08 &  ... & 0.21 \\
HD 147888     & $R(1,1)^u$ &  ... &   10 & ...  & 0.46 &  ... & 1.92 \\
              & $R(1,0)$   &  ... &   10 & ...  & 0.46 &  ... & 1.17 \\
HD 147889     & $R(1,1)^u$ &  ... &   10 & ...  & 0.18 &  ... & 0.74 \\
              & $R(1,0)$   &  ... &   10 & ...  & 0.18 &  ... & 0.45 \\
$\chi$ Oph    & $R(1,1)^u$ &  ... &    5 & ...  & 0.11 &  ... & 0.45 \\
              & $R(1,0)$   &  ... &    5 & ...  & 0.11 &  ... & 0.27 \\
$\mu$ Nor     & $R(1,1)^u$ &  ... &    5 & ...  & 0.12 &  ... & 0.49 \\
              & $R(1,0)$   &  ... &    5 & ...  & 0.12 &  ... & 0.30 \\
HD 149404     & $R(1,1)^u$ &  ... &    5 & ...  & 0.23 &  ... & 0.94 \\
              & $R(1,0)$   &  ... &    5 & ...  & 0.23 &  ... & 0.57 \\
$\zeta$ Oph   & $R(1,1)^u$ &  ... &   10 & ...  & 0.08 &  ... & 0.31 \\
              & $R(1,0)$   &  ... &   10 & ...  & 0.08 &  ... & 0.19 \\
HD 152236     & $R(1,1)^u$ &  ... &    5 & ...  & 0.14 &  ... & 0.59 \\
              & $R(1,0)$   &  ... &    5 & ...  & 0.14 &  ... & 0.36 \\
BD -14 5037   & $R(1,1)^u$ &  ... &   10 & ...  & 0.23 &  ... & 0.95 \\
              & $R(1,0)$   &  ... &   10 & ...  & 0.23 &  ... & 0.58 \\
HD 193322A    & $R(1,1)^u$ &  ... &   15 & ...  & 0.37 &  ... & 1.55 \\
              & $R(1,0)$   &  ... &   15 & ...  & 0.37 &  ... & 0.95 \\
Cyg OB2 8A    & $R(1,1)^u$ &  ... &   15 & ...  & 0.98 &  ... & 4.07 \\
              & $R(1,0)$   &  ... &   15 & ...  & 0.98 &  ... & 2.48 \\
HD 206267     & $R(1,1)^u$ &  ... &   10 & ...  & 0.30 &  ... & 1.25 \\
              & $R(1,0)$   &  ... &   10 & ...  & 0.30 &  ... & 0.76 \\
19 Cep        & $R(1,1)^u$ &  ... &   15 & ...  & 0.26 &  ... & 1.10 \\
              & $R(1,0)$   &  ... &   15 & ...  & 0.26 &  ... & 0.67 \\
1 Cas         & $R(1,1)^u$ &  ... &   15 & ...  & 0.29 &  ... & 1.21 \\
              & $R(1,0)$   &  ... &   15 & ...  & 0.29 &  ... & ~0.74
\label{table_lineparam}
\end{longtable}
%\end{landscape}
\normalsize

%%%%%%%%%%%%%%%%reaction rate coefficients%%%%%%%%%%%%%%%%%%%%%%%%%%%%%%%%%
\clearpage
\small
\begin{longtable}{llc}
\caption{Rate Coefficients for Reactions Involved in H$_3^+$ Chemistry}\\
Reaction & Rate Coefficient (cm$^3$ s$^{-1}$) & Reference \\
\hline
\endhead
\hline
\multicolumn{3}{p{6.0in}}{{\bf Notes:} Numerical subscripts on the rate coefficients refer to the corresponding reaction numbers given in the text. The H$_3^+$-electron recombination rate coefficient, $k_{\ref{re_H3+_e}}$, is referred to as $k_e$ throughout the text.  The rate coefficient for destruction via proton transfer to CO used in the text, $k_{\rm CO}$, is equal to $k_{\ref{re_H3+_CO_HCO+}}+k_{\ref{re_H3+_CO_HOC+}}$.  The rate coefficient for destruction via proton transfer to O, $k_{\ref{re_H3+_O}}$, is referred to as $k_{\rm O}$ throughout the text.  All temperature dependent coefficients are in terms of the gas kinetic temperature, $T$, except for $k_e$, which is in terms of the electron temperature, $T_e$.  However, it is expected that electrons and H$_2$ should be thermalized to the gas kinetic temperature via collisions, so we make no distinction between $T_e$, $T_{01}$, and $T$ in our calculations.   \newline
{\bf References:} (1) \citet{theard1974}; (2) \citet{mitchell1990}; (3) \citet{karpas1979}; (4) \citet{mccall2004}; (5) \citet{klippenstein2010}; (6) \citet{rakshit1982}
}
\endlastfoot
${\rm H}_2^+ + {\rm H}_2\rightarrow {\rm H}_3^+ + {\rm H}$ & $k_{\ref{re_H2_H2+}}=2.08\times10^{-9}$
& 1 \\
${\rm H}_2^+ + e^-\rightarrow \mathrm{H + H}$ & $k_{\ref{re_H2+_e}}=1.6\times10^{-8}(T/300)^{-0.43}$
& 2 \\
${\rm H}_2^+ + {\rm H}\rightarrow {\rm H}_2 + {\rm H}^+$ & $k_{\ref{re_H2+_H}}=6.4\times10^{-10}$
& 3 \\
${\rm H}_3^+ + e^-\rightarrow{\rm products}$ & $k_{\ref{re_H3+_e}}=k_e=-1.3\times10^{-8}+1.27\times10^{-6}T_e^{-0.48}$
& 4 \\
${\rm H}_3^+ + {\rm CO}\rightarrow\mathrm{H_2 + HCO^+}$ & $k_{\ref{re_H3+_CO_HCO+}}=1.36\times10^{-9}(T/300)^{-0.142}\exp(3.41/T)$
& 5 \\
${\rm H}_3^+ + {\rm CO}\rightarrow\mathrm{H_2 + HOC^+}$ & $k_{\ref{re_H3+_CO_HOC+}}=8.49\times10^{-10}(T/300)^{0.0661}\exp(-5.21/T)$
& 5 \\
${\rm H}_3^+ + {\rm O}\rightarrow{\rm H}_2 + {\rm OH}^+$  & $k_{\ref{re_H3+_O}}=k_{\rm O}=1.14\times10^{-9}(T/300)^{-0.156}\exp(-1.41/T)$        & 5 \\
${\rm H}_3^+ + {\rm N}_2\rightarrow{\rm H}_2 + {\rm HN}_2^+$ & $k_{\ref{re_H3+_N2}}=1.8\times10^{-9}$                                     & ~6
\label{table_H3+ratecoeff}
\end{longtable}
\normalsize

%%%%%%%%%%%%%%%%%%%%%%%%%%%%%%%%%%cosmic-ray ionization rate%%%%%%%%%%%%%%%%%%%%%%%%%%%%%%%%%%
\clearpage
\scriptsize
\begin{landscape}
\begin{longtable}{ccccccccccc}
\caption{Cosmic-Ray Ionization Rates}\\
       & $N({\rm H}_3^+)$      & $\sigma[N({\rm H}_3^+)]$ & H$_3^+$   & $N({\rm H}_2)$        & $\sigma[N({\rm H}_2)]$ & H$_2$     &
 $n_{\rm H}$ & Density   & $\zeta_2$            & $\sigma(\zeta_2)$     \\
Object & ($10^{13}$ cm$^{-2}$) & ($10^{13}$ cm$^{-2}$)    & Reference & ($10^{20}$ cm$^{-2}$) & ($10^{20}$ cm$^{-2}$)  & Reference &
 (cm$^{-3}$) & Reference &($10^{-16}$ s$^{-1}$) & ($10^{-16}$ s$^{-1}$) \\
\hline
\endfirsthead
\caption[]{(continued)}\\
       & $N({\rm H}_3^+)$      & $\sigma[N({\rm H}_3^+)]$ & H$_3^+$   & $N({\rm H}_2)$        & $\sigma[N({\rm H}_2)]$ & H$_2$     &
 $n_{\rm H}$ & Density   & $\zeta_2$            & $\sigma(\zeta_2)$     \\
Object & ($10^{13}$ cm$^{-2}$) & ($10^{13}$ cm$^{-2}$)    & Reference & ($10^{20}$ cm$^{-2}$) & ($10^{20}$ cm$^{-2}$)  & Reference &
 (cm$^{-3}$) & Reference &($10^{-16}$ s$^{-1}$) & ($10^{-16}$ s$^{-1}$) \\
\hline
\endhead
\hline
\multicolumn{11}{p{8.0in}}{\footnotesize{\bf Notes:} The sight lines toward WR 118 and HD 183143 have 2 distinct velocity components observed in H$_3^+$ by \citet{mccall2002}, and each of these components is labeled with a number in parentheses that corresponds to its average LSR velocity.  When no reference for $n_{\rm H}$ was available the value was set to 200 cm$^{-3}$, an average density for diffuse molecular clouds.  Upper limits to the cosmic-ray ionization rate should be taken as 3 times $\sigma(\zeta_2)$. \newline
{\bf H$_3^+$ References:} (1) this work; (2) \citet{mccall2002}; (3) \citet{crabtree2011}; (4) \citet{mccall1998} \newline
{\bf H$_2$ References:} (5) calculated from CH column densities
% (see Table \ref{table_CHandCHp})
using the relation $N({\rm CH})/N({\rm H}_2)=3.5^{+2.1}_{-1.4}\times10^{-8}$ from \citet{sheffer2008}; (6) \citet{savage1977}; (7) \citet{rachford2002}; (8) calculated from $E(B-V)$ (see Table \ref{table_targets}) using the relation $N_{\rm H}\approx E(B-V)5.8\times10^{21}$ cm$^{-2}$ mag$^{-1}$ from \citet{bohlin1978}, and assuming $2N({\rm H}_2)/N_{\rm H}=f_{{\rm H}_2}=0.67$; (9) \citet{rachford2009} \newline
$\mathbf{n_{\rm H}}$ {\bf References:} (10) \citet{sonnentrucker2007}; (11) \citet{shuping1999}; (12) \citet{jenkins1983}; (13) \citet{jura1975} }
\endlastfoot
HD 20041          & 23.7 & 5.13 & 1,2  & 11.4 & 6.24 & 5 & 200 &    & 9.48 & 7.59 \\
HD 21389          & 8.67 & 0.92 & 1    & 5.57 & 3.06 & 5 & 200 &    & 7.11 & 5.54 \\
$\zeta$ Per        & 6.26 & 0.52 & 1    & 4.75 & 0.95 & 6 & 215 & 10 & 5.55 & 3.18 \\
X Per                  & 7.34 & 0.92 & 1    & 8.38 & 0.89 & 7 & 325 & 10 & 5.85 & 3.54 \\
HD 29647          & 11.9 & 1.11 & 1    & 22.9 & 12.5 & 5 & 200 &    & 2.38 & 1.85 \\
HD 73882          & 9.02 & 0.50 & 3    & 12.9 & 2.39 & 7 & 520 & 10 & 9.71 & 5.57 \\
HD 110432        & 5.22 & 0.17 & 3    & 4.37 & 0.29 & 7 & 140 & 10 & 3.86 & 2.10 \\
HD 154368        & 9.37 & 1.32 & 1    & 14.4 & 3.99 & 7 & 240 & 10 & 4.19 & 2.62 \\
WR 104              & 23.2 & 1.54 & 2    & 44.7 & 22.3 & 8 & 200 &    & 2.37 & 1.76 \\
HD 168607        & 6.60 & 0.63 & 1    & 17.4 & 8.47 & 5 & 200 &    & 1.73 & 1.28 \\
HD 168625        & 10.6 & 0.92 & 1    & 16.3 & 7.95 & 5 & 200 &    & 2.96 & 2.17 \\
HD 169454        & 5.93 & 0.34 & 1    & 16.6 & 8.37 & 5 & 300 & 10 & 2.45 & 1.83 \\
W40 IRS 1A        & 26.9 & 3.54 & 1    & 56.3 & 28.2 & 8 & 300 & 11 & 3.27 & 2.45 \\
WR 118 (5)         & 29.4 & 7.07 & 2    & 43.3 & 21.7 & 8 & 200 &    & 3.10 & 2.41 \\
WR 118 (47)       & 30.8 & 5.73 & 2    & 45.5 & 22.7 & 8 & 200 &    & 3.10 & 2.36 \\
WR 121              & 22.0 & 2.60 & 2    & 34.0 & 17.0 & 8 & 200 &    & 2.96 & 2.21 \\
HD 183143 (7)   & 11.3 & 2.98 & 2    & 4.86 & 2.38 & 5 & 200 &    & 10.6 & 8.23 \\
HD 183143 (24) & 13.5 & 2.81 & 2    & 7.91 & 3.85 & 5 & 200 &    & 7.82 & 5.93 \\
HD 229059        & 37.4 & 1.20 & 1    & 65.7 & 51.2 & 5 & 200 &    & 2.60 & 2.47 \\
Cyg OB2 5          & 24.0 & 3.29 & 2    & 15.2 & 7.39 & 5 & 225 & 10 & 8.13 & 6.03 \\
Cyg OB2 12        & 34.3 & 5.89 & 2,4  & 80.0 & 69.1 & 5 & 300 & 10 & 2.93 & 3.04 \\
HD 204827        & 19.0 & 2.54 & 1    & 20.9 & 10.2 & 5 & 450 & 10 & 9.32 & 6.92 \\
$\lambda$ Cep  & 7.58 & 1.17 & 1    & 6.88 & 0.48 & 7 & 115 & 10 & 2.84 & 1.61 \\
$\kappa$ Cas    &  ... & 0.91 & 1    & 1.88 & 0.20 & 6 & 200 &    &  ... & 1.78 \\
HD 21483          &  ... & 2.30 & 1    & 11.9 & 5.91 & 5 & 200 &    &  ... & 0.88 \\
HD 21856          &  ... & 2.25 & 1    & 1.10 & 0.12 & 6 & 200 &    &  ... & 8.42 \\
40 Per                &  ... & 0.87 & 1    & 2.92 & 0.53 & 6 &  80 & 12 &  ... & 0.57 \\
$o$ Per              &  ... & 0.47 & 1    & 4.09 & 0.79 & 6 & 265 & 10 &  ... & 0.85 \\
BD +31 643       &  ... & 4.63 & 1    & 12.4 & 4.39 & 7 & 200 &    &  ... & 1.67 \\
$\epsilon$ Per   &  ... & 0.38 & 1    & 0.33 & 0.07 & 6 &  15 & 13 &  ... & 0.36 \\
$\xi$ Per           &  ... & 0.51 & 1    & 3.42 & 0.53 & 6 & 300 & 13 &  ... & 1.98 \\
62 Tau               &  ... & 1.79 & 1    & 6.23 & 1.12 & 7 & 280 & 10 &  ... & 2.06 \\
$\alpha$ Cam    &  ... & 1.48 & 1    & 2.17 & 0.26 & 6 & 150 & 13 &  ... & 2.17 \\
$\chi^2$ Ori      &  ... & 2.75 & 2    & 4.90 & 1.05 & 9 & 200 &    &  ... & 2.79 \\
HD 47129          &  ... & 2.91 & 1    & 3.51 & 0.58 & 6 & 200 &    &  ... & 3.81 \\
HD 53367          &  ... & 0.44 & 1    & 11.1 & 1.44 & 9 & 200 &    &  ... & 0.20 \\
$o$ Sco             &  ... & 0.41 & 1    & 19.1 & 9.81 & 5 & 225 & 10 &  ... & 0.11 \\
HD 147888        &  ... & 2.25 & 1    & 2.97 & 0.61 & 9 & 215 & 10 &  ... & 9.25 \\
HD 147889       &  ... & 0.87 & 1    & 28.8 & 14.0 & 5 & 525 & 10 &  ... & 0.36 \\
$\chi$ Oph       &  ... & 0.52 & 1    & 4.26 & 1.02 & 6 & 185 & 10 &  ... & 0.64 \\
$\mu$ Nor        &  ... & 0.58 & 1    & 2.77 & 0.49 & 6 & 200 &    &  ... & 1.09 \\
HD 149404       &  ... & 1.10 & 1    & 6.17 & 0.53 & 9 & 200 &    &  ... & 0.87 \\
$\zeta$ Oph     &  ... & 0.37 & 1,2  & 4.49 & 0.42 & 6 & 215 & 10 &  ... & 0.39 \\
HD 152236       &  ... & 0.69 & 1    & 5.34 & 1.32 & 9 & 200 &    &  ... & 0.63 \\
BD -14 5037     &  ... & 1.11 & 1    & 38.6 & 23.6 & 5 & 200 &    &  ... & 0.13 \\
P Cygni             &  ... & 1.84 & 2    & 3.14 & 1.61 & 5 & 200 &    &  ... & 2.67 \\
HD 193322A     &  ... & 1.82 & 1    & 1.21 & 0.18 & 6 & 200 &    &  ... & 6.52 \\
HD 194279       &  ... & 4.59 & 2    & 23.7 & 11.9 & 8 & 200 &    &  ... & 0.88 \\
Cyg OB2 8A      &  ... & 4.77 & 1    & 9.26 & 4.58 & 5 & 200 &    &  ... & 2.35 \\
HD 206267       &  ... & 1.47 & 1    & 7.18 & 0.59 & 7 & 200 &    &  ... & 0.97 \\
19 Cep              &  ... & 1.28 & 1    & 1.21 & 0.18 & 6 & 200 &    &  ... & 4.60 \\
1 Cas                 &  ... & 1.41 & 1    & 1.40 & 0.19 & 6 & 200 &    &  ... & 4.18 \\
NGC 2024 IRS 1 &  ... & 1.27 & 1    & 49.0 & 24.5 & 8 & 10000 &  &  ... & ~0.14
\label{table_zeta2}
\end{longtable}
\end{landscape}
\normalsize

%%%%%%%%%%%%%%%%%%%%%%%%%%%%%%%%%%%%%%%%figures%%%%%%%%%%%%%%%%%%%%%%%%%%%%%%%%%%%%%%%%%%%%%%

%%%%%%%%%%%%%%%%%%%%%SPECTRAAAAA!!!!!!!!!!!!!!!!!!!!!!!!!!!!!!!!%%%%%%%%%%%%%%%%%%%%%%%%%%%%%%%%%%%%%%%%%

\clearpage

\clearpage

\clearpage

\clearpage

\clearpage

\clearpage

\clearpage

\clearpage

\clearpage

\clearpage

\clearpage

\clearpage

\clearpage

%%%%%%%%%%%%%%%%%%%%%%%%%justify zeta equation%%%%%%%%%%%%%%%%%%%%%%%%
\clearpage

%%%%%%%%%%%%%%%%%%%%%%%%%%%zeta2 and mean%%%%%%%%%%%%%%%%%%%%%%%%%%%%%%%%
\clearpage

%%%%%%%%%%%%%%%%%%%%%%%%%%probability density function & other stats%%%%%%%%%%%%%%%%%%%%%%%%%%
\clearpage

%%%%%%%%%%%%%%%%%%%%%%%%%%%H3+ vs H2%%%%%%%%%%%%%%%%%%%%%%%%%%%%%%%%
\clearpage

%%%%%%%%%%%%%%%%%%%%%%%%%%%zeta2, distance, glat vs glong%%%%%%%%%%%%%%%%%%%%%%%%%%%%%%%%

%%%%%%%%%%%%%%%%%%%%%%%%%%%%%%%%zeta2 vs NH and E(B-V)%%%%%%%%%%%%%%%%%%%%%%%%%%%%%%%%%%%%%%%%%%%%%%%%

\end{document}